\documentclass[twocolumn,amsmath,amssymb,prd,superscriptaddress,nofootinbib,floatfix,showpacs]{revtex4}

\usepackage{graphicx}
\usepackage{color}
\usepackage{subfigure}
\usepackage{hyperref}
\usepackage{dcolumn}

\newcommand{\dslash}{\ensuremath{\partial\hspace{-1.2ex} /}}
\newcommand{\Tr}{\ensuremath{\operatorname{Tr}}}

\newcommand{\bj}{\marginpar{\colorbox{green}{\textbf{BJ}}\\@MW}}

\def\roughly#1{\mathrel{\raise.3ex\hbox{$#1$\kern-.75em%
\lower1ex\hbox{$\sim$}}}}

\newcommand\T{\rule{0pt}{2.6ex}}
\newcommand\B{\rule[-1.2ex]{0pt}{0pt}}

\newcommand{\ket}[1]{\left|{#1}\right\rangle}

\newcommand{\einh}[1]{\ensuremath{\,\text{#1}}}
\newcommand{\MeV}{\einh{MeV}}

\def\g2k{\Gamma^{(2)}_k}

\def\Eq#1{Eq.~(\ref{#1})}
\def\Fig#1{Fig.~\ref{#1}}
\def\Tab#1{Tab.~\ref{#1}}
\def\Eqs#1{Eqs.~(\ref{#1})}
\def\thetas{\theta_S}
\def\thetap{\theta_P}
\newcommand{\fpi}{\ensuremath{f_{\pi}}}
\newcommand{\fk}{\ensuremath{f_{K}}}
\newcommand{\mpi}{\ensuremath{m_{\pi}}}
\newcommand{\mk}{\ensuremath{m_{K}}}
\newcommand{\msig}{\ensuremath{m_{\sigma}}}
\newcommand{\metap}{\ensuremath{m_{\eta '}}}
\newcommand{\mkappa}{\ensuremath{m_{\kappa}}}
\newcommand{\meta}{\ensuremath{m_{\eta}}}
\def\ma0{m_{a_{0}}}
\def\mf0{m_{f_{0}}}
\newcommand{\ens}{{\eta_{\text{NS}}}}
\newcommand{\signs}{{\sigma_{\text{NS}}}}
\newcommand{\sigs}{{\sigma_{\text{S}}}}
\newcommand{\es}{{\eta_{\text{S}}}}

\newcommand{\ua}{\ensuremath{U(1)_A}}

\newcommand{\bsig}{\ensuremath{\bar{\sigma}}}
\newcommand{\mwc}[1]{\multicolumn{1}{c|}{#1}}
\newcommand{\mwmc}[2]{\multicolumn{1}{#1}{#2}}
\usepackage{xspace}
\newcommand{\lsm}{L\ensuremath{\sigma}M\xspace}
\newcommand{\multicite}[2]{\cite{#2}}

\graphicspath{
{./}
{./qm3_figures/}
{../figures/}
}
\begin{document}

\title{The three-flavor chiral phase structure in hot and dense QCD
  matter}

\author{B.-J. Schaefer}
\email[E-Mail:]{bernd-jochen.schaefer@uni-graz.at}
\affiliation{Institut f\"{u}r Physik, Karl-Franzens-Universit\"{a}t,
  A-8010 Graz, Austria}

\author{M. Wagner}
\email[E-Mail:]{mathias.wagner@physik.tu-darmstadt.de}
\affiliation{Institut f\"{u}r Kernphysik, TU Darmstadt, D-64289
  Darmstadt, Germany}

\date{\today}

\begin{abstract}
  Chiral symmetry restoration at nonzero temperature and quark
  densities are investigated in the framework of a linear sigma model
  with $N_f=3$ light quark flavors. After the derivation of the grand
  potential in mean-field approximation, the nonstrange and
  strange condensates, the in-medium masses of the scalar and
  pseudoscalar nonets are analyzed in hot and dense medium. The
  influence of the axial anomaly on the nonet masses and the isoscalar
  mixings on the pseudoscalar $\eta$-$\eta'$ and scalar
  $\sigma(600)$-$f_0(1370)$ complex are examined. The sensitivity of
  the chiral phase transition as well as the existence and location of
  a critical end point in the phase diagram on the value of the sigma
  mass is explored. The chiral critical surface with and without the
  influence of the axial $\ua$ anomaly is elaborated as a function of
  the pion and kaon masses for several values of the sigma mass.
\end{abstract}

\maketitle


\section{Introduction}

The understanding of the properties of strongly interacting matter
under extreme conditions is one of the most fascinating and
challenging tasks. General features of hot and dense matter are
summarized in the QCD phase diagram which can be probed by
ultrarelativistic heavy ion experiments such as the RHIC (BNL), LHC
(CERN) and the planned future CBM experiment at the FAIR facility in
Darmstadt.

Theoretical considerations indicate that at high temperature and high
baryon densities there should be a phase transition from ordinary
hadronic matter to a chirally symmetric plasma of quarks and gluons
\multicite{Svetitsky:1985ye, MeyerOrtmanns:1996ea,
  Rischke:2003mt}{mSvetitsky:1985ye}. Several issues concerning this
transition are not yet clarified \multicite{Stephanov:2007fk,
  Stephanov:2004wx}{mStephanov:2007fk}. QCD in this temperature and
density regime is a strongly coupled theory and hence perturbation
theory cannot be used. In the absence of a systematically improvable
and converging method to approach QCD at finite density one often
turns to model investigations see e.g.~\multicite{Klevansky:1992qe,
  Hatsuda:1994pi, Buballa:2003qv}{mKlevansky:1992qe}. These models
incorporate the important chiral symmetry breaking mechanism of QCD
but neglect any effects of confinement. Only recently, certain aspects
of confinement based on the Polyakov loop have been incorporated in
chiral effective models in a systematic fashion
\cite{Meisinger:1995ih, Fukushima:2003fw} and interesting conclusions
could be drawn (see e.g.~\multicite{Ratti:2005jh, Megias:2004hj,
  Schaefer:2007pw, Sasaki:2006ww, Abuki:2008nm,
  Fukushima:2008wg}{mRatti:2005jh}).

The most prominent finding from low-energy chiral effective models is
the QCD critical end point (CEP) \multicite{Asakawa:1989bq,
  Barducci:1989wi, Barducci1994, Stephanov:1998dy, Halasz:1998qr,
  Berges:1998rc}{mAsakawa:1989bq}. Common to almost all effective
model calculations is that the chiral phase transition is continuous
in the low density region and discontinuous in the high density
regime. Consequently, the endpoint where the phase transition ceases
to be discontinuous is the QCD critical end point. Unfortunately,
several obvious and related features such as the exact location of
this point in the QCD phase diagram cannot be predicted by these
models.

On the other side, lattice QCD simulations are important
alternatives to effective models calculations and can gain much
insights in the QCD phase structure \cite{Karsch:2001cy, Fodor:2002km,
  Allton2005gk, Aoki:2006br, Karsch:2004wd, Karsch:2007vw,
  Cheng:2006qk, Cheng:2007jq}. Due to the notorious
sign problem emerging at finite baryon density reliable predictions
for QCD are still not conclusive. Even worse, recently different
lattice methods that circumvent the sign problem are in conflict to
each other. For example, using the imaginary chemical potential method
for three physical quark masses no critical endpoint in the phase
diagram is found \cite{deForcrand:2003hx, D'Elia:2004at,
  D'Elia:2002gd, deForcrand:2006pv, deForcrand:2007rq,
  Philipsen:2007rj}.

The present work is an extension of a previous analysis within an
effective linear sigma model (\lsm) with two quark flavors to three
quark flavors \cite{Schaefer:2006ds, Schaefer:2004en}. The restoration
of the chiral $SU(3)\times SU(3)$ and axial $\ua$ symmetries with
temperature and quark chemical potentials are investigated. The axial
$\ua$ anomaly is considered via an effective 't~Hooft determinant in
the Lagrangian which breaks the $\ua$ symmetry. The restoration of the
$\ua$ symmetry is linked to a vanishing of the topological
susceptibility which can further be related to the $\eta'$ mass via
the Witten-Veneziano relation \multicite{Witten:1979vv, Veneziano:1979ec,
  Alkofer:2008et}{mWitten:1979vv, Alkofer:2008et}.

Some results depend sensitively on the model parameters which are
tuned to reproduce the vacuum phenomenology. There are model-input
parameters such as the $\sigma$ meson mass which are poorly known
experimentally. The generic findings of several parameter fits over a
broad range of input parameters are compared. Furthermore, the
extrapolation towards the chiral limit is also addressed and the mass
sensitivity of the chiral phase structure is investigated.

The paper is organized as follows: after introducing the \lsm with
three quark flavors, some symmetry breaking patterns in the vacuum are
briefly discussed. In Sec.~\ref{sec:potential} the grand
thermodynamic potential is derived in mean-field approximation. In
Sec.~\ref{sec:parameter_fits} a discussion of the model parameter fits
is given. Since the experimental situation concerning the scalar
$\sigma$-meson, $\sigma(600)$, is not settled we consider a
wide range of different values of the $\sigma$-meson mass, $\msig$, as
input parameter. All parameter sets are collected in
App.~\ref{app:parameters}.

For the \lsm without quarks it is known that the standard
loop expansion and related approximation methods at finite
temperatures fail and imaginary meson masses are generated. In our
approximation no such artifacts occur which enables us to perform a
careful and detailed analysis of chiral symmetry restoration in hot
and dense matter. This is demonstrated in Sec.~\ref{sec:finite_T}
where the pseudoscalar and scalar meson masses at finite temperatures
and chemical potentials with and without axial $\ua$ symmetry breaking
are investigated. All mass expressions are summarized in
App.~\ref{sec:meson_masses}. In addition, the scalar and pseudoscalar
flavor mixing behavior in the medium is explored. Various definitions
are deferred to App.~\ref{sec:eta_mixing}

The grand potential determines all thermodynamic properties.
The resulting phase diagrams are presented in Section
\ref{sec:crit_surface} where the mass sensitivity of the chiral phase
boundaries is also explored. Subsequently, the shape of the chiral
critical surface which confines the region of the chiral first-order
transitions in the $\mpi$-$m_K$ plane at the critical chemical
potential is evaluated for several values of $\msig$. Finally, in
Sec.~\ref{sec:summary} a summary with concluding remarks is given.

\section{Linear sigma model with three quark flavors}

The Lagrangian, ${\cal L}_{qm} = {\cal L}_q + {\cal L}_m$, of the
$SU(3)_L \times SU(3)_R$ symmetric \lsm with three quark
flavors consists of the fermionic part
\begin{equation}
\label{eq:quarkL}
\mathcal{L}_{q} = \bar{q}\left(i \dslash - g\; T_a\left(\sigma_a + i
    \gamma_5 \pi_a\right) \right) q 
\end{equation}
with a flavor-blind Yukawa coupling $g$ of the quarks to the mesons
and the purely mesonic contribution
\begin{eqnarray}
\label{eq:mesonL}
  \mathcal{L}_m &=& \Tr \left( \partial_\mu \phi^\dagger \partial^\mu
    \phi \right)
  - m^2 \Tr ( \phi^\dagger \phi) -\lambda_1 \left[\Tr (\phi^\dagger
    \phi)\right]^2 \nonumber \\
  &&  - \lambda_2 \Tr\left(\phi^\dagger \phi\right)^2
  +c   \left(\det (\phi) + \det (\phi^\dagger) \right)\nonumber \\
  && + \Tr\left[H(\phi+\phi^\dagger)\right]\ .
\end{eqnarray}
The column vector $q=(u, d, s)$ denotes the quark field for $N_f = 3$
flavors and $N_c=3$ color degrees of freedom \cite{Gell-Mann:1960np}.
The $\phi$-field represents a complex $(3\times 3)$-matrix and is
defined in terms of the scalar $\sigma_a$ and the pseudoscalar $\pi_a$
meson nonet
\begin{equation}
  \label{eq:fields}
  \phi = T_a \phi_a = T_a \left(\sigma_a + i \pi_a\right)\ .
\end{equation}
The $T_a = \lambda_a/2$ with $a=0,\ldots, 8$ are the nine generators
of the $U(3)$ symmetry, where the $\lambda_a$ are the usual eight
Gell-Mann matrices and $\lambda_0 = \sqrt{\frac{ 2}{3}}\ \bf 1$. The
generators $T_a$ are normalized to $\Tr (T_a T_b) = \delta_{ab}/2 $
and obey the $U(3)$ algebra $[T_a,T_b]=if_{abc}T_c$ and
$\{T_a,T_b \}=d_{abc}T_c$ respectively with the corresponding standard
symmetric $d_{abc}$ and antisymmetric $f_{abc}$ structure constants of
the $SU(3)$ group and
\begin{eqnarray}
f_{ab0} = 0 &,& d_{ab0} = \sqrt{\frac{ 2}{3}} \delta_{ab}\ .
\end{eqnarray}
Chiral symmetry is broken explicitly by the last term in
\Eq{eq:mesonL} where
\begin{equation}
  H = T_a h_a
\end{equation}
is a $(3\times 3)$-matrix with nine external parameters $h_a$. In
general, one could add further explicit symmetry breaking terms to
${\cal L}_{m}$ which are non-linear in $\phi$ \cite{mckay1973,
  Pagels:1974se} but this is ignored in this work.

Due to spontaneous chiral symmetry breaking in the vacuum a finite
vacuum expectation value of the $\phi$ field, $\bar{\phi}$, is
generated which must carry the quantum numbers of the vacuum
\cite{Gasiorowicz:1969kn}. As a consequence, only the diagonal
components $h_{0}, h_{3}$ and $h_{8}$ of the explicit symmetry
breaking term can be nonzero. This in turn involves three finite
condensates $\bsig_0$, $\bsig_3$ and $\bsig_8$ of which $\bsig_3$
breaks the $SU(2)$ isospin symmetry. In the following we shall
restrict ourselves to a $2+1$ flavor symmetry breaking pattern and
neglect the violation of the isospin symmetry. This is reflected by
the choice $h_{0} \neq 0, h_{3}=0, h_{8} \neq 0$ and corresponds to
two degenerated light quark flavors ($u,d$) and one heavier quark
flavor ($s$).

Besides the explicit symmetry breaking terms $h_0$ and $h_8$ the model
has five more parameters: the squared tree-level mass of the meson
fields $m^2$, two possible quartic coupling constants $\lambda_1$ and
$\lambda_2$, a Yukawa coupling $g$ and a cubic coupling constant $c$
which models the axial $\ua$ anomaly of the QCD vacuum. The $\ua$
symmetry of the classical QCD Lagrangian is anomalous
\cite{Weinberg:1975ui}, i.e.~broken by quantum effects. Without the
anomaly a ninth pseudoscalar Goldstone boson corresponding to the
spontaneous breaking of the chiral $U(3)_L\times U(3)_R$ symmetry
should emerge. However, experimentally, the lightest candidate for
this boson is the $\eta'$ meson, whose mass is of the order
$\metap \sim 960$ MeV which is far from being a light Goldstone boson.
The explicit breaking of the $\ua$ symmetry is held responsible for
the fact that the $\eta'$ mass is considerably larger than all other
pseudoscalar meson masses. This well-known $\ua$ problem of QCD is
effectively controlled by the anomaly term $c$ in the Lagrangian. The
comprehensive procedure of how to fix the parameters will be given in
Sec.~\ref{sec:parameter_fits}.

Depending on the signs and values of the parameters several
possible symmetry breaking patterns in the vacuum can be obtained (see
also \cite{Pagels:1974se} for more details). Without explicit symmetry
breaking terms, i.e.~for $H=0$, and without an explicit $\ua$ symmetry
breaking term, i.e.~for $c=0$, the Lagrangian has a global
$SU(3)_V\times U(3)_A \simeq SU(3)_V\times SU(3)_A \times \ua$
symmetry if the quartic coupling $\lambda_2$ and $m^2$ are
positive\footnote{More precisely, without the determinant the
  Lagrangian ${\cal L}_{qm}$ is $U(3 )\times U(3)$ invariant which is
  isomorphic to $SU(3)_L \times SU(3)_R \times U(1)_B \times \ua$. The
  $U(1)_B$ is related to the baryon number and always conserved which
  is why we have neglected it.}. If the mass parameter $m^2$ changes
sign the symmetry is spontaneously broken down to $SU(3)_V$. The
first quartic coupling $\lambda_1$ has no influence on the symmetry
breaking. Because of the breaking of the $U(3)_A$ symmetry nine
pseudoscalar Goldstone bosons arise which form the entire nonet
consisting of three pions, four kaons, the $\eta$ and $\eta'$ meson.
The scalar nonet belongs to the $SU(3)_V$ group which has a singlet
and an octet representation. All masses of the octet particles are
degenerate. The $\sigma$ meson belongs to the singlet and its mass is
in general different from the masses of the other octet particles.

By setting $c\neq 0$ the effects of the $\ua$ symmetry breaking, caused
by a nonvanishing topological susceptibility, are included and the
symmetry of the Lagrangian is reduced to $SU(3)_V\times SU(3)_A$. Due
to the spontaneous symmetry breaking of the $SU(3)_A$, the vacuum has
a $SU(3)_V$ symmetry. In this case the entire pseudoscalar octet is
degenerated and only eight Goldstone bosons appear. The $\eta'$ meson,
the would-be Goldstone boson, is still massive in this case. The
masses of the scalar particles are not modified by the $U(1)_A$
breaking.

With explicit symmetry breaking terms, more precisely, for only finite
$h_0$ and $h_8$ terms, the vacuum $SU(3)_V$ symmetry is explicitly
broken down to the isospin $SU(2)_V$ symmetry since the $h_3$ term is
set to zero. This symmetry pattern is already a good approximation to
nature because the violation of the isospin symmetry is small anyway.
The resulting ground state spectrum for this symmetry pattern will be
discussed in Sec.~\ref{sec:meson_spectrum}.

\section{Grand potential}
\label{sec:potential}

In this section, the derivation of the grand thermodynamic potential
for the three-flavor model is given. We will use a mean-field
approximation similar to the one for the two-flavor model in
\cite{Schaefer:2006ds}. The mean-field approximation is simple in its
application, in particular at finite temperature and quark densities.
Low-energy theorems, such as e.g. the Goldstone theorem or the Ward
identities are also fulfilled at finite temperatures and densities. In
this way we can circumvent more advanced many-body resummation
techniques which are usually necessary to cure the breakdown of naive
perturbation theory due to infrared divergences. For example, it is
well-known that the standard loop expansion or related expansion
methods of the $SU(3)$ version of the \lsm with or without quarks
break down at finite temperature and imaginary meson masses are
generated in the spontaneously broken phase \cite{Goldberg:1983ju,
  Meyer-Ortmanns:1992pj, Meyer-Ortmanns:1994nt, Lenaghan:2000ey,
  Roder:2003uz}. Contributions of thermal excitations to the meson
masses are neglected in these approximation schemes which result in a
too rapid decrease of the meson masses and the e.g.~squared pion mass
becomes negative for temperatures much below the phase transition.
This deficiency can be cured by self-consistent resummation schemes
such as e.g.~the Hartree approximation in the CJT formalism
\cite{Lenaghan:2000ey, SchaffnerBielich:1998zi} or the so-called
Optimized Perturbation Theory (OPT) e.g.~\cite{Chiku:1998kd} and
variants thereof. Recently, the OPT method has been frequently applied
to the three flavor \lsm with and without quarks at finite temperature
and baryon densities \multicite{Herpay:2005yr, Herpay:2006vc,
  Kovacs:2007sy, Kovacs:2006ym}{mHerpay:2005yr, Kovacs:2006ym}.
However, the predictive power of the OPT method depends on how it is
implemented and approximations thereof are made. For instance, when
the external momentum of the self-energy are taken on-shell a solution
of the corresponding gap equation and also of the equation of state
cease to exist above a certain temperature, particularly below the
critical one. Details and some improvements of certain approximations
in the OPT framework can be found in \multicite{Herpay:2005yr,
  Herpay:2006vc, Kovacs:2007sy, Kovacs:2006ym}{mHerpay:2005yr,
  Kovacs:2006ym}.

All these problems do not emerge in the mean-field approximation used
here. This enables us to study the phase structure of the more
involved three flavor model in great detail and in a rather simple
framework.

In order to calculate the grand potential in mean-field
approximation we start from the partition function. In thermal
equilibrium, the grand partition function is defined by a path
integral over the quark/antiquark and meson fields
\begin{equation}
  \label{eq:partition}
  \mathcal{Z}\!=\!\!\! \int\! \prod_a \mathcal{D}\sigma_a \mathcal{D}\pi_a \!\!\int\!\!
  \mathcal{D}q \mathcal{D}\bar{q} \exp \left( - \int_0^{1/T}
    \!\!\!\!\!\!d\tau \!\!\int_V \!\!\!d^3x
    \mathcal{L}^E \right),
\end{equation}
where $T$ is the temperature and $V$ the three-dimensional volume of
the system\footnote{An irrelevant normalization
  constant is suppressed.}. For three quark flavors the Euclidean
Lagrangian $\mathcal{L}^E$ generally contains  three independent quark
chemical potentials $\mu_f$
\begin{equation*}
  \mathcal{L}^E = \mathcal{L}_{qm} + \sum_{f=u,d,s} \mu_f q^\dagger_f
  q_f\ .
\end{equation*}
Due to the assumed $SU(2)_V$ isospin symmetry we neglect the slight
mass difference between an $u$- and $d$-quark and the light quark
chemical potentials become equal. In the following we denote the
degenerated light quark quantities by an index $q$, i.e.~the light
quark chemical potential by $\mu_q \equiv \mu_u = \mu_d$, and the
strange quark quantities by an index $s$.

The calculation of the partition function in the mean-field
approximation is performed similar to Refs.~\cite{Scavenius:2000qd,
  Schaefer:2006ds} for the two-flavor case. The quantum and thermal
fluctuations of the mesons are neglected and the quarks/antiquarks are
retained as quantum fields. This means that the integration over the
mesonic fields in \Eq{eq:partition} is dropped and the fields are
replaced by their non-vanishing vacuum expectation values
$\bar \phi = T_{0} \bsig_{0} + T_{8} \bsig_{8}$. The remaining
integration over the Grassmann fields yields a determinant which can
be rewritten as a trace over a logarithm. Evaluating the trace within
the Matsubara formalism, the quark contribution
$\Omega_{\bar{q}q} (T,\mu_f)$ of the grand potential is
obtained \cite{Kapusta1989}. The ultraviolet divergent vacuum
contribution to $\Omega_{\bar{q}q} (T,\mu_f)$ which results from the
negative energy states of the Dirac sea has been neglected here
(cf.~\cite{Scavenius:2000qd, Schaefer:2006ds} for further details).
Finally, the total grand potential is obtained as a sum of the quark
contribution and meson contribution $U(\bsig_0, \bsig_8)$ as

\begin{equation}
  \label{eq:grand_pot}
  \Omega(T,\mu_f) = \frac{-T \ln \mathcal{Z}}{V} = U \left( \bsig_0,
    \bsig_8 \right) + \Omega_{\bar{q}q}(T,\mu_f)\ .
\end{equation}
Explicitly, the quark contribution reads
\begin{multline}
  \label{eq:quark_pot}
  \Omega_{\bar{q}q}(T,\mu_f) = \nu_c T \sum_{f=u,d,s}
  \int\limits_0^\infty \! \frac{d^3 k}{(2\pi)^3}
  \left\{ \ln (1-n_{q,f}(T,\mu_f))\right.\\
  \left. + \ln (1-n_{\bar{q},f}(T,\mu_f)) \right\}
\end{multline}
with the usual fermionic occupation numbers for the quarks
\begin{equation}
  n_{q,f}\left(T,\mu_{f}\right)=\frac{1}{1+\exp\left((E_{q,f}-\mu_f)/T\right)}
\end{equation}
and antiquarks $n_{\bar q,f}(T,\mu_{f}) \equiv n_{q,f} (T,-\mu_{f})$
respectively. The number of internal quark degrees of freedom is
denoted by $\nu_c=2 N_{c} = 6$. The flavor-dependent single-particle
energies are
\begin{equation}
  E_{q,f}= \sqrt{k^2 + m_f^2}
\end{equation}
with the flavor-dependent quark masses $m_f$ which are also functions
of the expectation values $\bsig_0$ and $\bsig_8$.

The vacuum condensates $\bsig_0$ and $\bsig_8$ are members of the
scalar ($J^P = 0^+$) nonet and both contain strange and non-strange
components. For the further analysis it is more convenient to convert
the condensates into a pure non-strange and strange part. This is
achieved by an orthogonal basis transformation from the
original octet-singlet basis ($\sigma_0, \sigma_8$) to the non-strange
($\sigma_x$) and strange ($\sigma_y$) quark flavor basis
\begin{equation}
\begin{pmatrix}
\sigma_x\\
\sigma_y
\end{pmatrix} = \frac{ 1}{\sqrt{3}}
\begin{pmatrix}
\sqrt{2}&1\\
1& -\sqrt{2}
\end{pmatrix}
\begin{pmatrix}
\sigma_0\\
\sigma_8
\end{pmatrix}\ .
\label{eq:xytrafo}
\end{equation}

As a consequence, the light quark sector decouples from the strange
quark sector (cf. e.g.~\cite{Kovacs:2006ym}) and the quark masses
simplify in this new basis to
\begin{align}\label{eq:qmasses}
  m_q = g \sigma_x /2 &\quad,\quad m_s = g \sigma_y /\sqrt{2}\ .
\end{align}
The meson potential modifies accordingly
\begin{multline}\label{eq:umeson}
  U(\sigma_{x},\sigma_{y}) = \frac{m^{2}}{2}\left(\sigma_{x}^{2} +
  \sigma_{y}^{2}\right) -h_{x} \sigma_{x} -h_{y} \sigma_{y}
 - \frac{c}{2 \sqrt{2}} \sigma_{x}^2 \sigma_{y}
\\
  + \frac{\lambda_{1}}{2} \sigma_{x}^{2} \sigma_{y}^{2}+
  \frac{1}{8}\left(2 \lambda_{1} +
    \lambda_{2}\right)\sigma_{x}^{4}+\frac{1}{8}\left(2 \lambda_{1} +
    2\lambda_{2}\right) \sigma_{y}^{4}\ ,
\end{multline}
therein the explicit symmetry breaking parameters $h_0$ and $h_8$ have
also been transformed according to \Eq{eq:xytrafo}. The order
parameters for the chiral phase transition are identified here with
the expectation value $\bsig_x$ for the non-strange and with $\bsig_y$
for the strange sector. They are obtained by minimizing the total
thermodynamic potential (\ref{eq:grand_pot}) in the non-strange and
strange directions
\begin{equation}
  \label{eq:eom}
  \left.\frac{ \partial \Omega}{\partial
      \sigma_x} = \frac{ \partial \Omega}{\partial \sigma_y}
  \right|_{\sigma_x=\bsig_x, \sigma_y=\bsig_y} = 0\ .
\end{equation}
The solutions of these coupled equations determine the behavior of the
chiral order parameters as a function of $T$ and chemical potentials,
$\mu_q$ and $\mu_s$. Note, that the in-medium condensates are also
labeled with a bar over the corresponding fields.

\section{Parameter fits}
\label{sec:parameter_fits}

The \lsm with three quark flavors has altogether seven
parameters $m^2$, $\lambda_1$, $\lambda_2$, $c$, $g$, $h_0$, $h_8$ and
two unknown condensates $\bsig_x$ and $\bsig_y$. The six parameters
$m^2$, $\lambda_1$, $\lambda_2$, $c$, $h_x$ and $h_y$ of the mesonic
potential are fixed in the vacuum by six experimentally known
quantities. Similar to Ref.~\cite{Lenaghan:2000ey},  we have chosen as input
the low-lying pseudoscalar mass spectrum, $\mpi$ and $m_K$, the
average squared mass of the $\eta$ and $\eta '$ mesons,
$\meta^2 + \metap^2$, and the decay constants of the pion and kaon,
$\fpi$ and $f_K$, and in addition the scalar $\sigma$ meson mass
$\msig$. We can then predict the scalar meson masses $\ma0$,
$\mkappa$, $\mf0$, the difference of the $\eta$,$\eta'$ squared
masses, $\meta^2 - \metap^2$ and the scalar and pseudoscalar mixing
angles $\thetas$, $\thetap$.

In analogy to e.g.~Ref.~\cite{Lenaghan:2000ey} the values of the
condensates are determined from the pion and kaon decay constants by
means of the partially conserved axial-vector current relation (PCAC).
In the strange--non-strange basis they are given by
\begin{equation}
  \label{eq:pcac}
  \bsig_{x} = f_{\pi} \  ; \qquad
  \bsig_{y} = \frac{1}{\sqrt{2}} \left(2 \fk - f_{\pi}\right)\ .
\end{equation}

The average squared $\eta$ and $\eta'$ meson mass determines the
parameter $\lambda_2$ by
\begin{widetext}
\begin{equation}
  \lambda_2 = \frac{3(2\fk-\fpi )\mk^2 - (2\fk+\fpi)\mpi^2
    -2(\metap^2+\meta^2)(\fk-\fpi)}{\left(3\fpi^2+8\fk
  (\fk-\fpi)\right)(\fk-\fpi) }\ ,
\end{equation}
\end{widetext}
and the $U(1)_A$ anomaly breaking term $c$ is fixed by $\lambda_2$ and
the difference of the pion and kaon masses squared via
\begin{equation}
  c = \frac{\mk^2-\mpi^2}{\fk-\fpi}-\lambda_2 (2 \fk - \fpi)   \ .
\end{equation}

Note, that without anomaly breaking, i.e.~for $c=0$, the average
$\eta$-$\eta'$ meson mass is not used anymore for fixing the parameter
$\lambda_2$. It is then given by the kaon and pion masses and decay
constants only,
\begin{equation}
\lambda_{2}=\frac{m_K^2-m_{\pi}^2}{(2 f_K-f_{\pi})} (f_{K}-f_{\pi})\ .
\end{equation}

The input parameters from the pseudoscalar sector involve only a
relation between $\lambda_1$ and $m^2$. Therefore, further input from
the scalar sector is necessary. In principle two possible options are
available. At first, the parameter $m^2$ is expressed as a function of
the yet undetermined parameter $\lambda_1$. This can be achieved by
using the equation for the pion mass or the kaon mass (see
App.~\ref{sec:meson_masses}). In this way the $m^2$ dependence of the
$\sigma$ meson mass (or of the $f_0(1370)$ meson mass) can be
transformed in a $\lambda_1$ dependence since the scalar mixing angle
does not depend on $m^2$. By fixing the mass of the $\sigma$ meson (or
of the $f_0(1370)$ meson) $\lambda_1$ is determined by solving the
corresponding equation. Afterwards, the $m^2$ parameter follows
immediately, since $\lambda_1$ is fixed.

The explicit symmetry breaking terms $h_{x}$ and $h_{y}$ in the
non-strange--strange basis are related to the pion and kaon masses by
the Ward identities
\begin{equation}
  \label{eq:wardids}
  h_{x} = f_{\pi} \mpi^2 \  ; \qquad
  h_{y} = \sqrt{2} f_{K} m_K^2 - \frac{\fpi \mpi^2}{\sqrt{2}}\ .
\end{equation}
These relations can be derived by using the gap equations,
\Eqs{eq:eom}. The last open parameter, the value of the Yukawa
coupling $g$, is fixed from the non-strange constituent quark mass
\begin{equation}
  g = 2 m_q / \bsig_x\ .
\end{equation}
For example, using for the light constituent quark mass a value of
$m_q = 300$ MeV we obtain $g \sim 6.5$ and can predict a strange
constituent quark mass $m_s \approx 433$ MeV.

Since the experimental situation concerning the broad $\sigma$ (or
$f_0 (600)$) resonance is not yet clear, cf.~\cite{Yao:2006px}, we
will use different input values for $\msig$ in the range of
$\msig = 400 - 1000$ MeV and will investigate its mass dependence on
various quantities (see also \cite{Caprini:2005zr}). In
App.~\ref{app:parameters} several parameter sets for different $\msig$
values with and without effects of the axial $\ua$ anomaly are
summarized (\Tab{tab:psets}). Furthermore, a discussion of the
parameter sets with respect to spontaneous symmetry breaking can be
found in this appendix. The corresponding predictions of the scalar
and pseudoscalar meson masses and mixing angles are collected in
\Tab{tab:massesvac}

\section{Chiral symmetry restoration}
\label{sec:finite_T}

Having fixed the model parameters we can now evaluate the grand
potential numerically. In the following we present our
results for the chiral symmetry restoration at finite temperature and
finite quark density with and without axial anomaly breaking.
Throughout this section, the axial anomaly breaking term is kept
constant, in particular, independent of the temperature and the
chemical potentials.

\subsection{Condensates}

The chiral phase structure of the underlying three-flavor model is
completely governed by the total thermodynamic potential. Hence, the
solution of the gap equations (\ref{eq:eom}) determines the behavior
of the condensates as a function of temperature and quark chemical
potentials. In general, the three quark chemical potentials are
independent but here we will consider symmetric quark matter and
define a uniform chemical potential $\mu \equiv \mu_q = \mu_s$.

\begin{figure}
  \includegraphics[width=\linewidth]{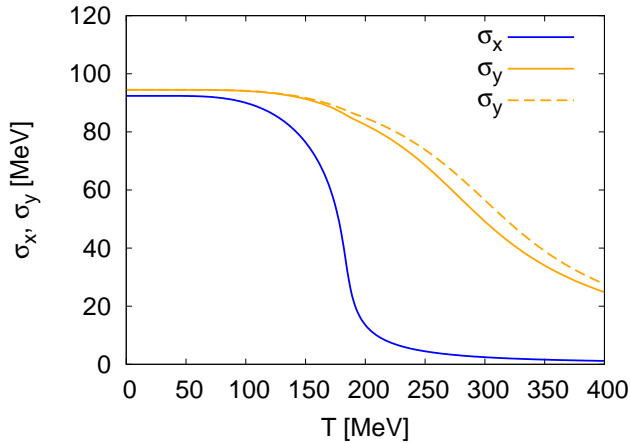}
  \caption{
    The non-strange, $\bsig_x$ and strange, $\bsig_y$ condensates as a
    function of temperature for vanishing chemical potentials with
    (solid) and without $U(1)_A$ anomaly (dashed). The anomaly does
    not modify the non-strange condensate.}
    \label{fig:condensates_mu0}
\end{figure}

In \Fig{fig:condensates_mu0} the nonstrange $\bsig_x$ and strange
$\bsig_y$ condensates are shown as a function of temperature for
vanishing chemical potential $\mu$ for $\msig = 800$ MeV. The reason
for choosing this value for $\msig$ will become clear later on, see
also App.~\ref{app:parameters}. The solid lines in this figure are
obtained with an explicit axial $\ua$ symmetry breaking term while the
dashed line corresponds to the anomaly free case, i.e. $c=0$. The
difference in the nonstrange condensate $\bsig_x$, caused by the
anomaly, is not visible in the figure. The condensates start at $T=0$
with the fitted values, $\bsig_x = 92.4$ MeV and $\bsig_y = 94.5$ MeV.
The temperature behavior of both condensates shows a smooth crossover.
The temperature derivative of the nonstrange condensate peaks around
$T\sim 181$ MeV. The precise value of this pseudo-critical temperature
depends on the value of $\msig$ in the vacuum. For smaller values of
$\msig$ the pseudocritical temperature decreases
(cf.~Sec.~\ref{sec:crit_surface}). The chiral transition in the
strange sector is much smoother due to the larger constituent strange
quark mass, $m_s = 433$ MeV. As a consequence, the chiral
$SU(2)\times SU(2)$ symmetry is restored more rapidly. With axial
anomaly the strange condensate melts a little earlier but only for
temperatures above the transition. At very high temperatures both
condensates become degenerate, indicating chiral $SU(3)\times SU(3)$
symmetry restoration.

If one uses a temperature dependent anomaly term by making use of
lattice results for the topological susceptibility which yields e.g.~a
decreasing anomaly term for increasing temperatures, a faster
effective restoration of the axial symmetry can be achieved, see
e.g.~\cite{SchaffnerBielich:1999uj, Costa:2005cz}.

\begin{figure}
  \includegraphics[width=\linewidth]{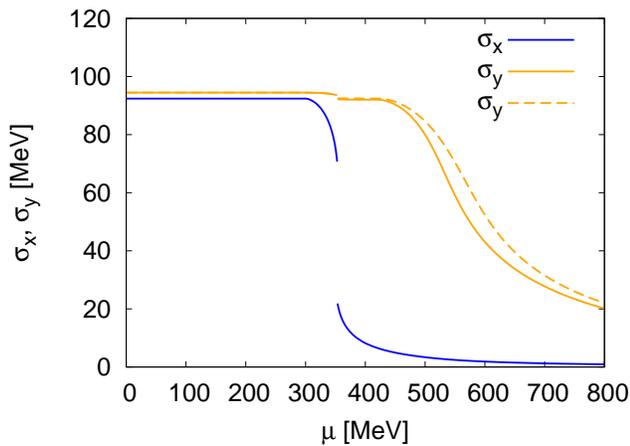}
  \caption{Similar to Fig.\ref{fig:condensates_mu0} but as a function
    of $\mu$ for $T=0$.}
   \label{fig:condensates_T5}
\end{figure}

For zero temperature and finite chemical potential both condensates
are independent of $\mu$ in the broken phase until the Fermi surface
of the light quarks is reached. For zero temperature the light Fermi
surface coincides with the light quark mass, i.e.~$\mu = m_q = 300$
MeV. Before the chiral transition takes place at a critical chemical
potential $\mu_c \sim 352$ MeV the nonstrange condensate drops by
about 10\% from its vacuum value as can be seen in
\Fig{fig:condensates_T5}. At $\mu_c$ the phase transition is of
first-order and three solutions of each gap equation (\ref{eq:eom})
appear corresponding to two degenerate minima and one maximum of the
effective potential. Due to the explicit symmetry breaking both
condensates remain always finite in the symmetric phase. The phase
transition is mainly driven by the nonstrange condensate while the
jump in the strange condensate is negligible. Above the transition for
$\mu > \mu_c$ and below the strange Fermi surface at
$\mu = m_s \sim 433$ MeV the strange condensate stays constant. The
axial anomaly has almost no influence up to the strange quark Fermi
surface. Only for chemical potentials larger than $\mu \sim 433$ MeV
the strange condensate melts faster if the $\ua$ symmetry breaking is
taken into account. For large chemical potentials this difference
vanishes again and both strange condensates will become identical.
Furthermore, the in-medium behavior of the nonstrange condensate
$\bar \sigma_x$ is not modified by the anomaly.

\begin{figure*}
  \subfigure[{$\ $} with {$\ua$} symmetry breaking
  breaking]{\label{sfig:pset1003_massestl_chiral1}
    \includegraphics[width=0.45\linewidth]{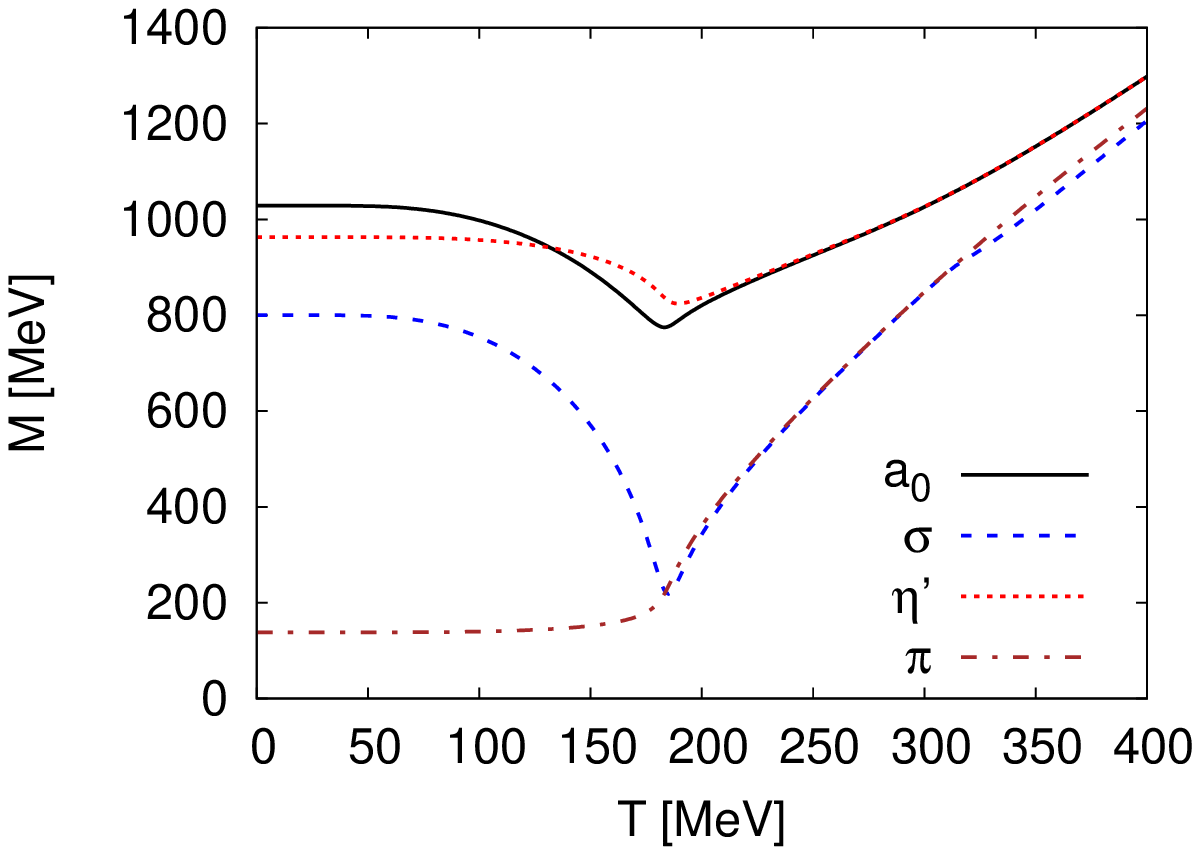}}
  \subfigure[{$\ $} without {$\ua$} symmetry breaking
  breaking]{\label{sfig:pset3_massestl_chiral1}
    \includegraphics[width=0.45\linewidth]{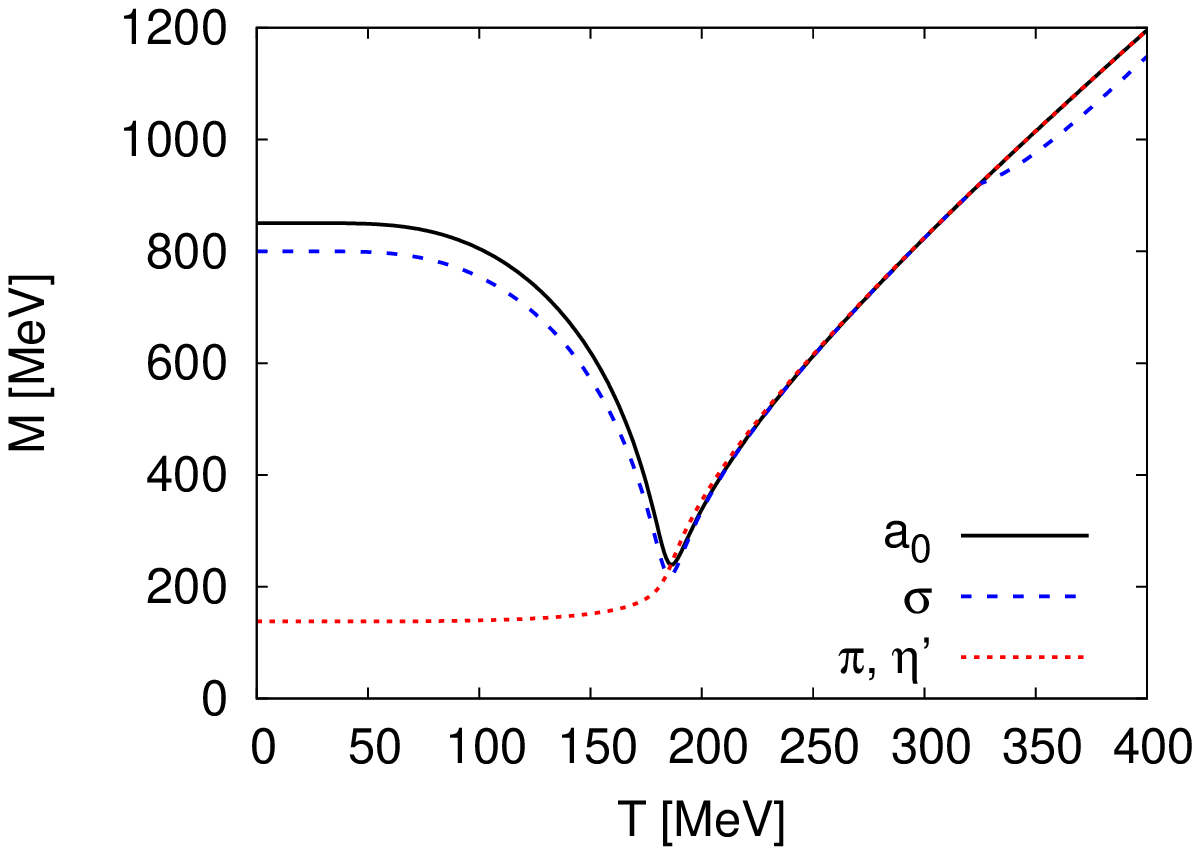}}
  \caption{In-medium meson masses, ($\pi$, $\sigma$) and ($\eta'$,
    $a_0$), as a function of temperature for $\mu=0$ with $U(1)_A$
    anomaly breaking (left panel). Without anomaly breaking (right
    panel) the $\eta'$ meson degenerates with the pion mass. See text
    for further details.}
  \label{fig:masses_chiral1}
\end{figure*}
\begin{figure*}
  \subfigure[{$\ $} with {$\ua$} symmetry breaking]{\label{sfig:pset1003_massestl_chiral2}
    \includegraphics[width=0.45\linewidth]{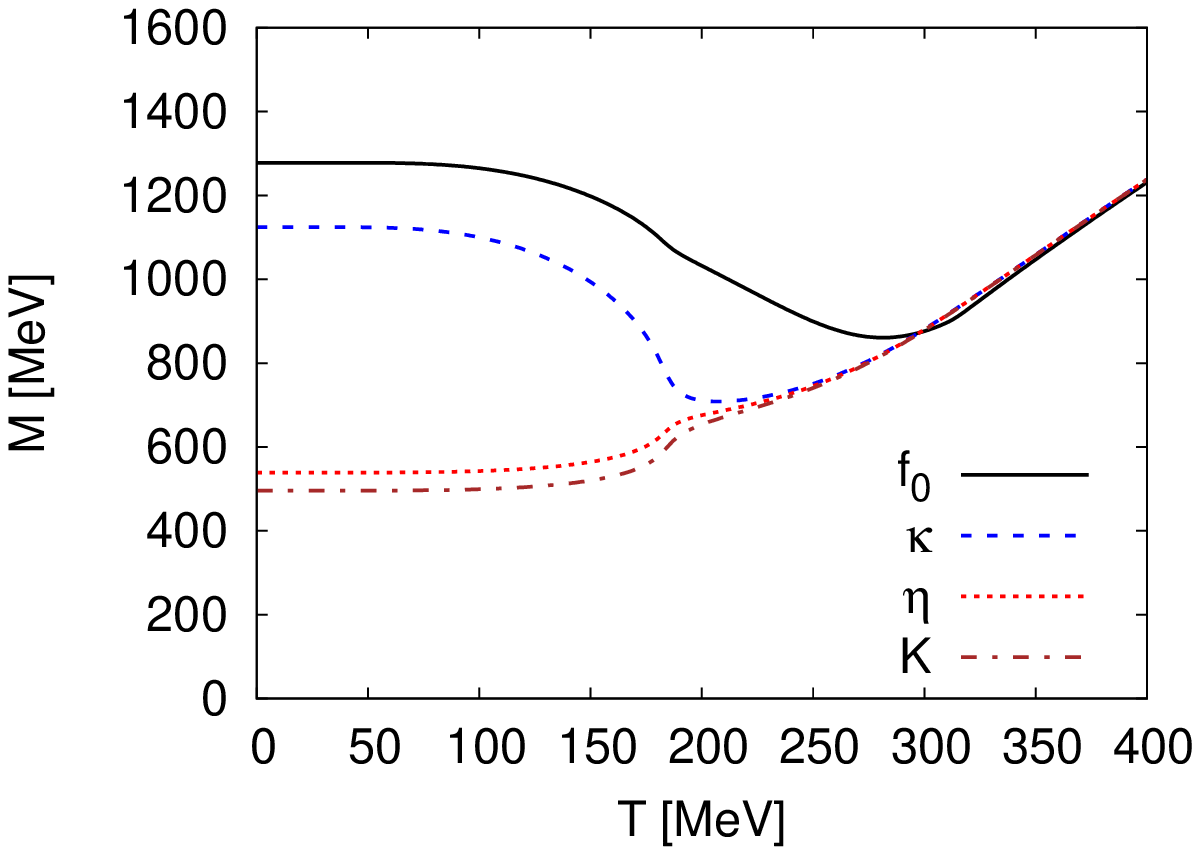}}
  \subfigure[{$\ $} without {$\ua$} symmetry
  breaking]{\label{sfig:pset3_massestl_chiral2}
    \includegraphics[width=0.45\linewidth]{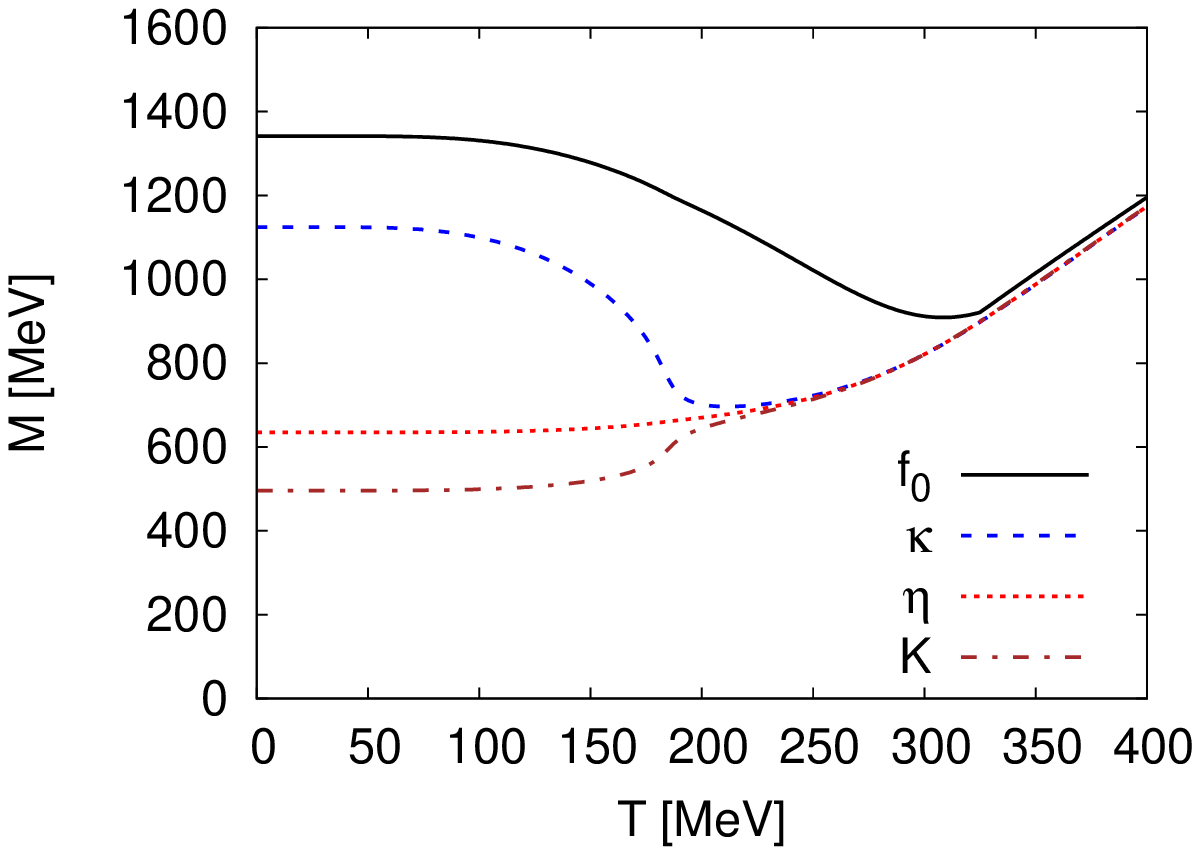}}
  \caption{Similar to Fig.~\ref{fig:masses_chiral1} for the chiral
    partners ($\eta$, $f_0$) and ($K$, $\kappa$) (left panel with
    $U(1)_A$ anomaly breaking). Without anomaly breaking (right panel)
    the $\eta$ meson increases about $100$ MeV in the vacuum.}
  \label{fig:masses_chiral2}
\end{figure*}

\subsection{The scalar-pseudoscalar meson spectrum}
\label{sec:meson_spectrum}

In the following the in-medium scalar and pseudoscalar meson mass
spectrum is analyzed. The derivation of the in-medium masses as well
as the mass formulae are collected in App.~\ref{sec:meson_masses}.

We start with the discussion of the mass spectrum at nonzero
temperature and vanishing quark chemical potential.
The meson masses as a function of the temperature for $\mu=0$ are
shown in Figs~\ref{fig:masses_chiral1} and \ref{fig:masses_chiral2}.
In the left panels of the respective figures the $\ua$ symmetry
breaking is explicitly taken into account while in the right panels
the breaking is neglected.

The masses of the pion and the $\sigma$ meson and also the masses of
the $\eta'$ and the $a_0$ meson degenerate approximately at the same
temperature $T_c \sim 181$ MeV. This temperature behavior signals the
effective restoration of chiral $SU(2)\times SU(2)$ symmetry in the
non-strange sector via a smooth crossover transition. The chiral
partners $(K, \kappa)$ show a similar temperature behavior but
degenerate with the $\eta$ meson at a higher temperature $T \sim 240$
MeV. At the chiral transition $T \sim 181$ MeV the $\kappa$ meson
becomes lighter than the $a_0$ meson. In contrast to
\cite{Lenaghan:2000ey} the $\eta'$ is always heavier than the kaon for
all temperatures. Only the $f_0(1370)$ meson mass does not show a
tendency to converge to the $\eta$ mass in the temperature
region shown because chiral symmetry in the strange sector is restored only
very slowly. The intersection point of the $f_0$ and the $\eta$ mass
coincides with the inflection point of the strange condensate.
Nevertheless, the $f_0$ meson will degenerate with the remaining meson
octet at very large temperatures.

The mass gap in the restored phase for $T>T_c$ between the two sets of
the chiral partners, $(\sigma, \pi)$ and $(a_0, \eta')$, i.e.
$\mpi=\msig < \ma0 = m_{\eta'}$ is a consequence of the $U(1)_A$
breaking term. This gap is generated by an opposite sign of the
anomaly term in the scalar and pseudoscalar meson masses,
cf.~App.~\ref{sec:meson_masses}. It is basically given by $\sqrt{2}c
\bsig_y$, i.e. proportional to the anomaly term $c$ and the strange
order parameter $\bsig_y$. The non-strange condensate $\bsig_x$ is
already negligible for temperatures above $T_c$. For higher
temperatures ($T \gg 400$ MeV) the $U(1)_A$ symmetry gets effectively
restored and the mass gap between the chiral partners will vanish.
Finally, for very large temperatures compared to the strange quark
mass the difference between the strange and non-strange mesons
disappear and all meson masses will degenerate.

Without $\ua$ symmetry breaking the mass gap between the chiral
partners, $(\sigma, \pi)$ and $(a_0, \eta')$ vanishes in the restored
phase and all four meson masses degenerate at the same critical
temperature $T_c \sim 181$ MeV coinciding with the inflection point of
the non-strange condensate. Above this temperature the axial symmetry
is restored but the full restoration of the $U(3)\times U(3)$ symmetry
is still not yet completed because the chiral partners $(K, \kappa)$
degenerate with the $\eta$ at a higher temperature $T\sim 240 \MeV$.
This temperature value and the value of $T_c$ are not changed by the
anomaly as expected since the non-strange condensate is not influenced
by the anomaly. Interestingly, a recent mean-field study within the
three-flavor NJL model with various effective $\ua$ anomaly
implementations found an explicit difference for the chiral
non-strange transition temperatures with and without explicit $\ua$
symmetry breaking (cf.~Tab.~III in \cite{Costa:2005cz}).

As for the case with anomaly, the chiral partners ($\eta, f_0$)
degenerate but only for temperatures around $300$ MeV because
these mesons are purely strange states and chiral symmetry in the
strange sector is very slowly restored. A mild anomaly dependence of
the intersection point of the $f_0$ and the $\eta$ meson is observed.
There is no inverse mass ordering of the $\eta$ meson and the kaon at
finite temperature as found in \cite{Lenaghan:2000ey}. In the vacuum
the mass of the $f_0$ increases by about $60$ MeV if the anomaly is
neglected.

Without the anomaly term the $\eta'$ meson degenerates with the pion
already in the vacuum and stays degenerated with the pion for all
temperatures. Hence, in the vacuum the mass of the $\eta'$ drops down
considerably from $963$ MeV to $138$ MeV. In fact, it has been shown
that the mass of the $\eta'$ must be less then $\sqrt{3}\mpi \sim 240$
MeV if the $\ua$ symmetry is not explicitly broken
\cite{Weinberg:1975ui}.

In general, one can summarize the mass spectrum in-medium behavior in
the following way: the bosonic thermal contributions decrease the
meson masses while the fermionic parts increase the masses. For small
temperatures the quark contribution is negligible and for high
temperatures it dominates the mesonic contribution yielding rising and
degenerate meson masses.

All meson masses are controlled by the two explicit symmetry breaking
parameter $h_x$ and $h_y$. They are determined by the tree-level Ward
identities, \Eq{eq:wardids}, which guarantee the Goldstone theorem at
zero temperature: for vanishing external parameter $h_x$ the pion mass
must also vanishes because $\fpi$ is then finite. In this case, the
other symmetry breaking parameter $h_y$ generates only a finite value
for the kaon mass.
Furthermore, the chiral limit can be reached by setting all explicit
symmetry breaking parameters to zero. But in order to obtain finite
vacuum expectation values for the condensates the symmetry must be
spontaneously broken. This requires a negative $m^2$ parameter.
Later, we will use several parameter fits for various values of the
sigma meson mass which partly have a positive $m^2$ parameter (see
App.~\ref{app:parameters}). For these parameter sets one cannot reach
the chiral limit by just setting the explicit symmetry breaking
parameters to zero. But these parameter sets are still well suited
for fitting the physical mass point. For instance, choosing a
$\msig =400$ MeV the parameter fit results with or without anomaly in
a positive $m^2$ parameter and the chiral limit cannot be reached for
this parameter set. In \cite{Lenaghan:2000ey} another strategy to
investigate the chiral limit was adopted by performing a separate extra
parameter fit where an average of the experimental mass values in the
scalar octet spectrum together with some extrapolated quantities
towards the chiral limit as input have been used. However, all in all
the extrapolation towards the chiral limit remains questionable for
both procedures.

For the parameter set with e.g.~$\msig=800$ MeV we can reach the
chiral limit. For $c\neq0$ we obtain a massless pseudoscalar octet and
a finite $\metap = 767$ MeV due to the $\ua$ symmetry breaking. All
scalar octet masses are degenerate at $840$ MeV and the mass of the
sigma meson drops down to $620$ MeV. Moreover, without $\ua$ symmetry
breaking all nine pseudoscalar mesons are massless and the scalar
octet masses are degenerated into $780$ MeV and $\msig = 712$ MeV.

In our approximation the Goldstone's theorem is also valid at finite
temperature and chemical potentials, meaning that in the chiral limit
the masses of the Goldstone bosons stay massless in the broken phase.
Even in the presence of quarks both Ward identities in
(\ref{eq:wardids}) are always fulfilled for all temperatures and quark
chemical potentials which can be shown analytically.

Another observation in the Figs~\ref{fig:masses_chiral1} and
\ref{fig:masses_chiral2} is the temperature behavior of the scalar
$\sigma$ and $f_0$ meson around $T \sim 325$ MeV. There is a kink
visible in the curves and the meson masses seem to interchange their
identities for higher temperatures. In order to elucidate this
behavior we analyze the scalar and pseudoscalar mixing angles in
the following.

\subsection{Flavor mixing at finite temperature}

\begin{figure}
  \includegraphics[width=\linewidth]{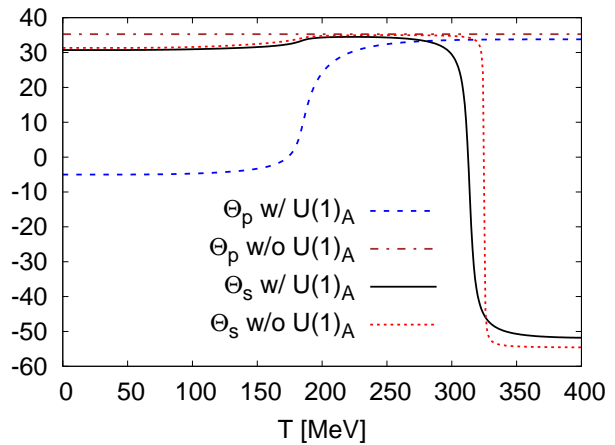}
  \caption{The scalar $\thetas$ and pseudoscalar $\thetap$ mixing
    angles as a function of temperature for $\mu=0$ without and with
    $U_A (1)$ anomaly.}
  \label{fig:pset3_1003_angles}
\end{figure}

The investigation of the mixing angles of the scalar and pseudoscalar
isoscalar states provides further insights of the axial $\ua$ symmetry
restoration. In order to clarify our findings some underlying
definitions and relations between different bases are given in
App.~\ref{sec:eta_mixing}. Both mixing angles, the pseudoscalar
$\thetap$ and the scalar one $\thetas$, are shown in
Fig.~\ref{fig:pset3_1003_angles} as a function of temperature for
$\mu=0$ with and without explicit $\ua$ symmetry breaking. In the
broken phase, i.e.~for $T<200$ MeV a strong influence of the anomaly
on the pseudoscalar sector is found while almost no effect is seen in
the scalar sector. With anomaly the nonstrange and strange quark
states mix and generate an pseudoscalar mixing angle
$\thetap \sim -5^\circ$ at $T=0$. For increasing temperatures the
mixing angle stays almost constant in the chirally broken phase.
Around the chiral restoration temperature $T_c \sim 180$ MeV the angle
increases smoothly towards the ideal mixing angle
$\thetap \to \arctan 1/\sqrt{2} \sim +35^\circ$ corresponding to
$\phi_p = 90^\circ$, where $\phi_p$ denotes the pseudoscalar mixing
angle in the strange-nonstrange basis (see App.~\ref{sec:eta_mixing}
for details). At high temperatures this means that the $\eta$ meson
becomes a purely strange and the $\eta'$ a purely non-strange quark
system (cf.~\multicite{Kunihiro:1989my, Alkofer:1989rr}{mKunihiro:1989my}).

\begin{figure}
  \includegraphics[width=\linewidth]{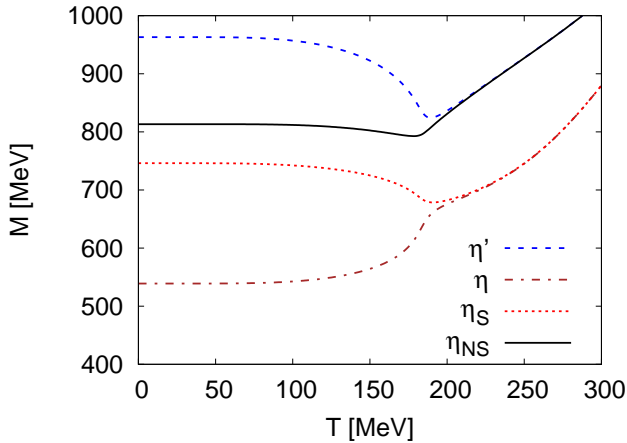}
  \caption{Physical $\eta$-$\eta'$ complex in comparison with the
    $\ens$-$\es$ complex as a function of temperature at $\mu=0$ with
    anomaly breaking.}
   \label{fig:pset3massesangles}
\end{figure}

This is also demonstrated in \Fig{fig:pset3massesangles} where the
physical $\eta$-$\eta'$ and the nonstrange-strange $\ens$-$\es$
complex are shown as a function of temperature for $\mu=0$. At $T=0$
the nonstrange mass of the $\ens$ meson, $m_\ens$, is larger than the
strange mass $m_\es$ since the pseudoscalar mixing angle is larger
than $\phi_p = 45^\circ$, respectively $\thetap = -9.74^\circ$. For
the mixing angle of $\thetap \sim -5^\circ$ ($\phi_p \sim
49.74^\circ$) we obtain $m_\ens \sim 813$ MeV and $m_\es \sim 746$
MeV.

At the chiral transition temperature $T_c \sim 180$ MeV the $\eta'$
meson becomes purely nonstrange ($\eta' \to \ens$) while the $\eta$
becomes a purely strange quark system ($\eta \to \es$). In this
temperature region the mixing angle grows to the ideal
$\thetap \to +35^\circ$ (respectively $\phi_p \to 90^\circ$). No
crossing of the $\ens$ and $\es$ or anticrossing\footnote{A crossing
  of the $\ens$ and $\es$ masses corresponds to an anticrossing of the
  physical $\eta$ and $\eta'$ masses via
  \Eqs{eq:etadiag}-(\ref{eq:etadiag2}) (cf.~also
  \cite{Horvatic:2007qs}).} of the physical $\eta$-$\eta'$ complex is
observed for all temperatures since $\phi_p (T)$ is always above
$45^\circ$ \multicite{Kunihiro:1989my, Alkofer:1989rr}{mKunihiro:1989my}.

Without anomaly the pseudoscalar mixing angle is already ideal for
zero temperature and stays ideal for all temperatures,
i.e.~$\thetap \sim +35^\circ$. This means that already at $T=0$ the
$\eta$ and $\eta'$ mesons are ideally flavor-mixed. The $\eta'$ is a
purely light quark system and the $\eta$ is a purely strange quark
system. Without anomaly the $\eta'$ degenerates in mass with the pion.
Hence, the $\eta'$ belongs to the class of nonstrange particles. The
ordering of the corresponding nonstrange-strange masses is reversed,
i.e. without anomaly $m_\es$ is larger than $m_\ens$ since
$m_\es = \meta$ and $m_\ens = \metap$.

\begin{figure}
  \includegraphics[width=\linewidth]{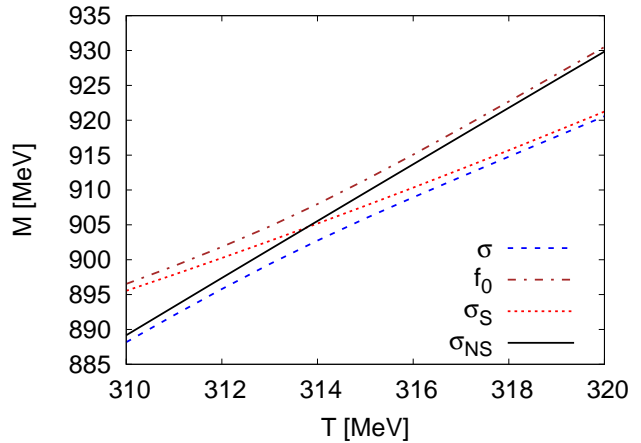}
  \caption{Similar to \Fig{fig:pset3massesangles}: The scalar physical
    $\sigma$-$f_0$ complex around $T\sim 314$ MeV in comparison with
    the $\signs$-$\sigs$ system with anomaly breaking.}
  \label{fig:pset1003_sigmamasses}
\end{figure}

In the scalar sector the mixing angle $\thetas$ shows no influence of
the axial anomaly in the broken phase (see
\Fig{fig:pset3_1003_angles}). In both cases, with and without anomaly
breaking, the mixing angle is almost ideal $\thetas \sim +31^\circ$ at
$T=0$ but the precise vacuum value depends strongly on the
value for the fitted scalar sigma meson mass in contrast to the
pseudoscalar angle which is independent of $\msig$, see
\Tab{tab:massesvac} in App.~\ref{app:parameters}. For increasing
$\msig$ the scalar mixing angle $\thetas$ also increases at $T=0$. As
a consequence, for larger $\msig$ the nonstrange-strange
$\signs$-$\sigs$ complex and the physical $\sigma$ and $f_0(1370)$
mesons degenerate more and more at $T=0$, meaning that the $\sigma$
meson tends to a pure nonstrange quark system, $\sigma \to \signs$,
and the $f_0$ to a pure strange system, $f_0 \to \sigs$. For instance,
we obtain for $\msig (m_{f_0}) = 400 (1257)$ MeV respectively
$m_\signs (m_\sigs) = 561 (1131)$ MeV and for
$\msig (m_{f_0}) = 800 (1341)$ MeV $m_\signs (m_\sigs) = 804 (1276)$ MeV.

At the chiral transition $\thetas$ grows again to the ideal one. But
for temperatures around $T\sim 314$ MeV in the chirally symmetric
phase the scalar mixing angle drops down to $\thetas \sim -54^\circ$
($\phi_s \sim 0^\circ$). Around these temperature the masses of the
physical $\sigma$ and $f_0$ anticross and the ones of the
nonstrange-strange $\signs$-$\sigs$ system cross. This is displayed
in \Fig{fig:pset1003_sigmamasses}. Hence, for larger temperatures,
$T>314$ MeV the $f_0$ is now a purely nonstrange quark system and the
$\sigma$ a purely strange state. For very large temperatures, around
900 MeV, the scalar mixing angle turns back to the ideal
$\thetas \sim +35^\circ$ again and a crossing and anticrossing of the
corresponding masses takes place again. Without anomaly the same
phenomenon happens qualitatively around some larger temperatures of
the order of $T\sim 325$ MeV.

For finite quark chemical potential and vanishing temperature the
mixing angle show qualitatively a similar behavior. Around
$\mu \sim 350$ MeV the pseudoscalar angle $\thetap$ increases towards
the ideal value while without anomaly the angle is already ideal. In
the scalar sector the angle is nearly ideal in the broken phase and
drops down to $\thetas \sim -54^\circ$ around $\mu \sim 500$ MeV where
again the masses of the physical $\sigma$ and $f_0$ meson anticross.

A finite temperature study of the $\eta$-$\eta'$ complex including the
QCD axial anomaly within a Dyson-Schwinger approach and a
temperature-dependent topological susceptibility can also be found in
\multicite{Klabucar:1997zi, Kekez:2005ie, Horvatic:2007qs,
  Klabucar:2001gr}{mKlabucar:1997zi, Horvatic:2007qs}. By means of the Witten-Veneziano relation the
authors studied the interplay between the melting of the topological
susceptibility and the chiral restoration temperature. The authors
find a strong increase of the $\eta'$ mass around the chiral
restoration temperature which makes the extension of the
Witten-Veneziano relation to finite temperature questionable. In the
present work a constant anomaly parameter has been used corresponding
to a constant topological susceptibility. This means that the $\ua$
symmetry is not restored around the chiral critical temperature but at
higher temperatures.

\section{Phase diagram and the chiral critical surface}
\label{sec:crit_surface}

The phase diagram is constructed in the following way: for realistic
pion and kaon masses (the so-called physical point) the light
condensate melts always faster with $T$ and/or $\mu$ than the strange
condensate because the strange quark mass is heavier than the light
quark mass. As the chiral phase boundary we use the inflection point
in the light condensate.

Later on, we will also vary the meson masses and calculate the
corresponding phase diagrams. As a consequence, the ordering of the
light and strange condensates can be inverted since the kaon mass can
become lighter than the pion mass. In such cases the strange
condensate drops faster than the light condensate and the chiral phase
transition is triggered by the strange condensate. This has to be
taken into account systematically, in particular, for the calculation
of the chiral critical surface. The faster melting condensate has been
used in order to localize the phase boundary.

\begin{figure}[htbp]
    \includegraphics[width=\linewidth]{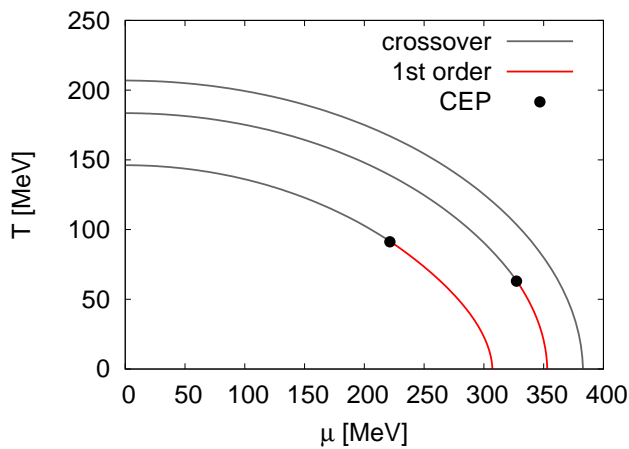}
    \caption{Phase diagrams with $\ua$ symmetry breaking for different values
      of $m_{\sigma}=600 \MeV$ (lower line), $800\MeV$, $900\MeV$
      (upper line).}
  \label{fig:phasediag_muB}
\end{figure}

In Fig.~\ref{fig:phasediag_muB} the phase diagrams in the $(T,
\mu)$-plane with explicit $\ua$ symmetry breaking for three different
values of $m_\sigma$ are shown (lower lines correspond to
$m_\sigma = 600$ MeV, next lines to $m_\sigma=800$ MeV and upper lines
to $m_\sigma = 900$ MeV). For each value of $\msig$ the remaining
parameters of the model are fitted to the vacuum as described in
Sec.~\ref{sec:parameter_fits} and are kept constant. Since the
explicit $\ua$ symmetry breaking leads only to small modifications of
the phase boundaries (cf.~Figs.~\ref{fig:condensates_mu0} and
\ref{fig:condensates_T5}), we refrain from presenting the phase diagrams in
the absence of an explicit $\ua$ symmetry breaking term.

At zero chemical potential a crossover is
always found due to the explicit symmetry breaking terms. The
crossover temperature depends on the choice of the sigma meson mass. For
increasing $\msig$ the pseudocritical temperature also increases (e.g.~for
$\msig = 600, 800, 900$ MeV we found a $T_c \sim 146, 184, 207$ MeV).

Recent lattice simulations at $\mu=0$ for three quark flavors have
obtained values for the pseudocritical temperatures in the range of
$T_c = 151(3)(3)$ MeV \cite{Karsch:2000kv, Cheng:2006qk} and
$T_c=192(7)(4)$ MeV \cite{Aoki:2006we,Aoki:2006br}. These results,
applied to the current study, would suggest values for the sigma mass
in between $600$ and $800$ MeV.

At zero temperature a first-order phase transition is obtained
(cf.~\Fig{fig:condensates_T5}). For increasing temperatures the
first-order transition becomes weaker and terminates in the critical
end point (CEP). How to measure this point and what the distinctive
signatures of this point are is not yet settled. It is interesting to
see that the mass of the $\sigma$ meson as a function of temperature
and/or chemical potential through the CEP always drops below the mass
of the pion not only for the corresponding two flavor but also for the
three flavor calculation \cite{Schaefer:2006ds}. This is a general
feature of the \lsm since the potential flattens at this point in
radial $\sigma$ direction. In a similar NJL calculation this is not
the case \cite{Scavenius:2000qd, costa-2008-77}.

For $\msig=600$ MeV the location of the CEP is at $(T_c, \mu_c) = (91,
221)$ MeV and for $\msig=800$ MeV at $(63,327)$ MeV. As a consequence
of the model parameters dependence, the location of the CEP moves for
increasing $\msig$ towards the $\mu$ axis. It is interesting to
observe that already for $\msig = 900$ MeV the existence of the CEP
disappears and the phase transition is a smooth crossover over the
entire phase diagram.

Almost no influence of the axial anomaly on the phase boundaries and
thus on the location of the CEP is seen. For comparison, with and
without the $\ua$ symmetry breaking and each for $\msig = 600$ MeV the
location of the CEP changes from $(T_c, \mu_c) = (91, 221)$ MeV to
$(89, 228)$ MeV. For $\msig =800$ MeV the changes are even smaller
$(T_c, \mu_c) = (63, 327)$ MeV to $(63,328)$ MeV.

In Ref.~\cite{Struber:2007bm} a gauged linear sigma model with chiral
$U(N_f)\times U(N_f)$ symmetry without quarks within the 2PI
resummation scheme has been considered. For the two flavor case and
neglecting the influence of the vector mesons the opposite behavior of
the chiral phase transition as a function of $\msig$ is observed: for
$\mu=0$ a crossover is seen for a small $\sigma$ mass ($\msig = 441$
MeV) and a first-order transition for a large $\sigma$ mass
($\msig = 1370$ MeV). If the vector mesons are included the transition
leads to a more rapid crossover and brings one closer to the
second-order critical point. The conclusion is that the critical endpoint
moves closer to the temperature axis. Thus, the inclusion of vector
mesons should improve the agreement with lattice QCD results since
usually, in chiral models the critical endpoint is located at smaller
temperatures and larger chemical potentials.

\begin{figure}[htbp]
  \includegraphics[width=\linewidth]{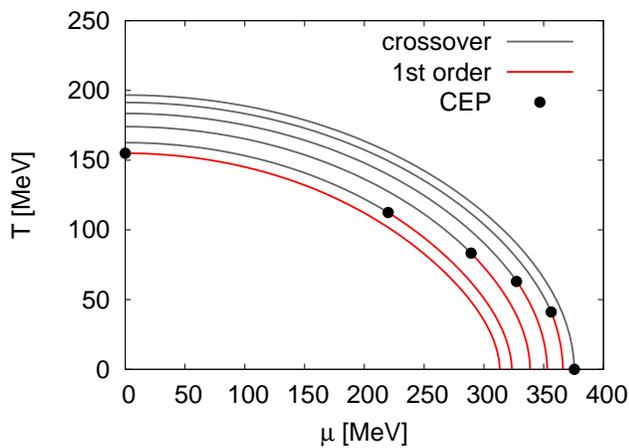}
  \caption{Phase diagrams with $\ua$ symmetry breaking for
    $\msig =800$ MeV and different pion masses: $\mpi/\mpi^*= 0.488$
    (lower line), $0.6, 0.8, 1.0, 1.2, 1.36$ (upper line),
    $m_{\pi}^{*}=138\MeV$, $m_K^{*}=496\MeV$. The ratio
    $m_{\pi}/m_{K}=m_{\pi}^{*}/m_{K}^{*}$ is kept fixed.}
  \label{fig:phasediag_muB_mpiK}
\end{figure}

For three quark flavors renormalization-group arguments predict a
first-order transition in the chiral limit independent of the $\ua$
symmetry breaking~\cite{Pisarski:1983ms}. This behavior is displayed
in Fig.~\ref{fig:phasediag_muB_mpiK} where the resulting phase
diagrams including the anomaly for varying pion and kaon masses are
shown for $\msig=800\MeV$. We have chosen a path in the $(\mpi,
\mk)$-plane through the physical mass point towards the chiral limit
by varying the pion mass and accordingly the kaon mass by keeping the
ratio $\mpi/\mk$ fixed at the value given at the physical point
$\mpi^*/\mk^*$. 

\begin{figure*}[th]
  \subfigure[{$\ $} with {$\ua$}
  anomaly]{\label{sfig:columbia_pset1003}
    \includegraphics[width=0.45\linewidth]{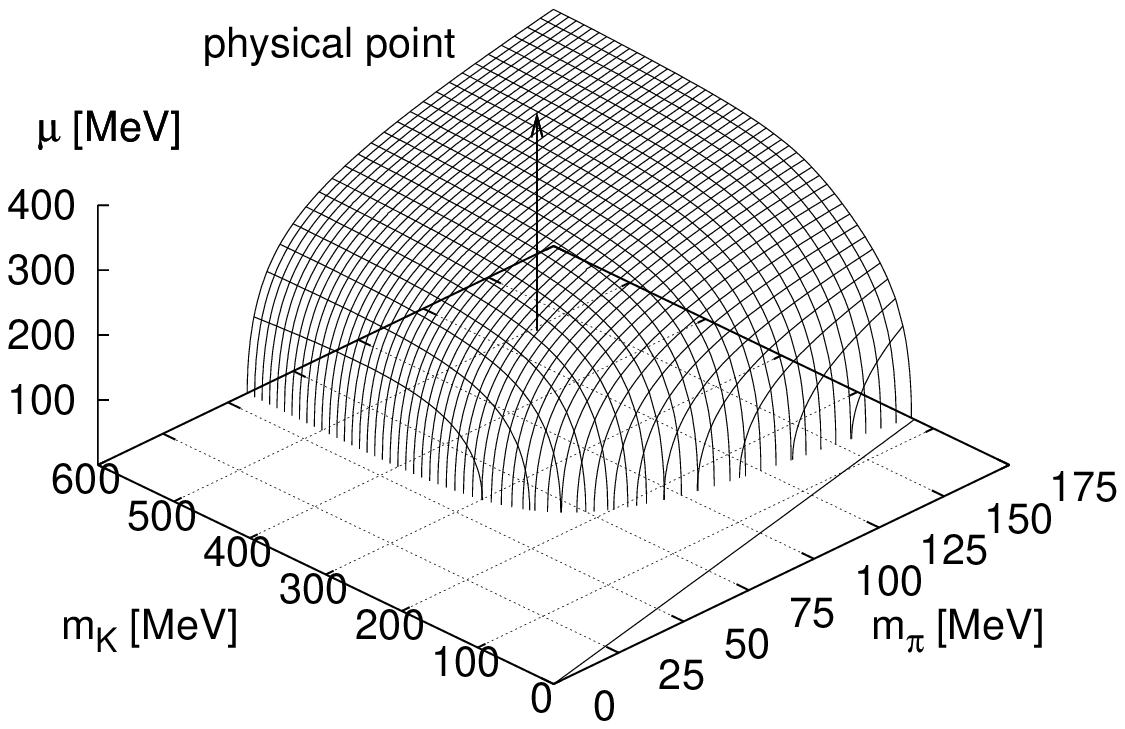}}
  \subfigure[{$\ $} without {$\ua$}
  anomaly]{\label{sfig:columbia_pset3}
    \includegraphics[width=0.45\linewidth]{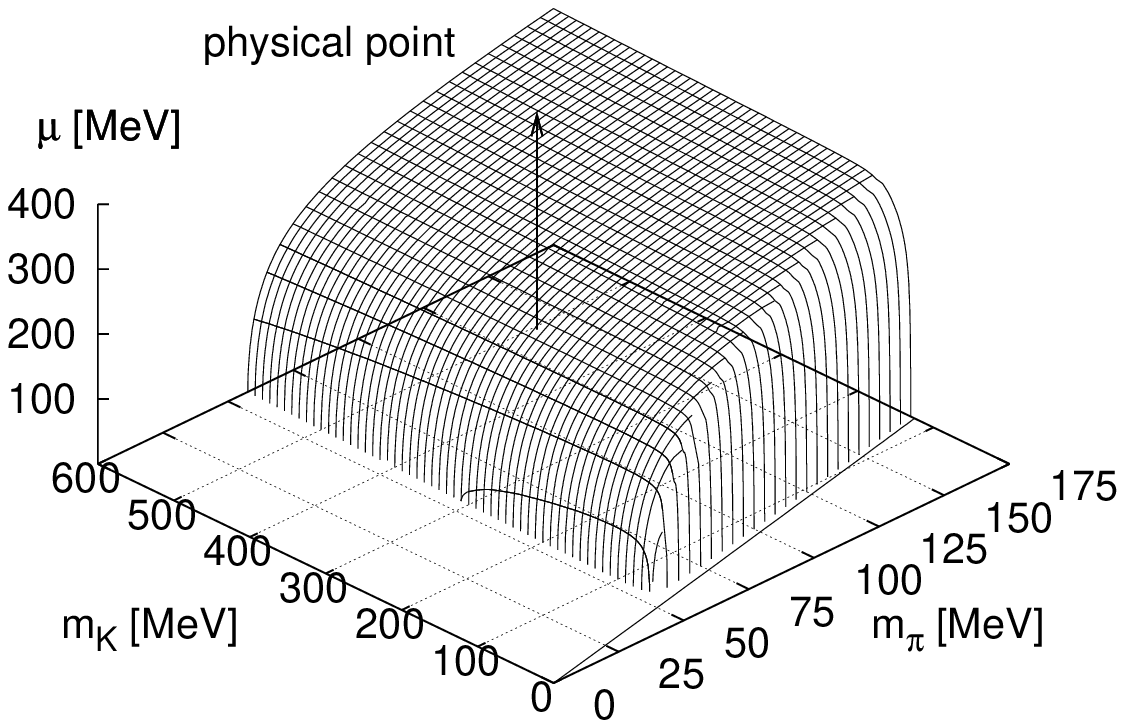}}
  \caption{The chiral critical surface in the ($m_\pi,m_K$) plane for
    $\msig = 800$ MeV. The arrow points to the critical quark chemical
    potential at realistic pion and kaon masses (physical point). The
    solid line in the right panel is given by $m_K =
    \mpi\sqrt{\fpi/2f_K}$. }
	\label{fig:columbia_pset3_1003}
\end{figure*}

The pion and kaon masses are varied by changing the explicit symmetry
breaking parameters, $h_x$ and $h_y$, while keeping all remaining
model parameters fixed at the values obtained at the physical point.
For a pion mass $1.36$ times larger than the physical one (for
$\msig = 800$ MeV) the CEP lies exactly on the $\mu$-axis, hence for
pion masses above this value the phase transition is a smooth
crossover over the entire phase diagram and no CEP exists anymore.

For a decreasing pion mass, the location of the CEP moves towards the
$T$-axis and already for half of the physical pion mass the CEP hits
the $T$-axis at $\mu=0$. Remarkably also for
half of the physical pion mass the CEP hits the $T$-axis in a similar
calculation within a two-flavor \lsm within the same
approximation scheme but without $\ua$ symmetry breaking and for
$\msig = 600$ MeV \cite{Schaefer:2006ds}. For smaller pion masses the
chiral phase transition turns into a first-order one for all densities
and consequently no CEP exists any longer.

The various transition lines in the phase diagram shrink towards the
origin of the phase diagram for smaller pion and kaon masses because
the condensates decrease more rapidly as a function of the temperature
and chemical potentials. If one rescales the temperature with the
critical temperature at $\mu=0$ and the chemical potential with the
critical chemical potential at $T=0$ then all the transition lines in
the phase diagram lie on top of each other for different pion masses.

In connection with the existence of the CEP in the phase diagram it is
interesting to analyze its mass sensitivity. For this purpose the
critical surface for the chiral phase transition is evaluated in
\Fig{fig:columbia_pset3_1003} as a function of the pion and the kaon
masses with (left panel) and without (right panel) the $\ua$ symmetry
breaking. Again the masses are tuned by variation of only the explicit
symmetry breaking parameters (see \Eq{eq:wardids}) while keeping all
other model parameters fixed similar to Ref.~\cite{Lenaghan:2000kr}.
For small kaon masses but large pion masses the explicit symmetry
breaking parameter for the strange direction $h_y$ in \Eq{eq:wardids}
can become negative. The corresponding kaon mass, where this happens,
is given by $\mk = \mpi \sqrt{\fpi/2\fk}$ and is shown in both panels
as a solid line.

The critical surface is defined by the value of the critical chemical
potential $\mu_c$ of the CEP for a given mass pair $(\mpi,m_K)$. It is
the surface of the second-order phase transition points displayed in a
three-dimensional $(\mu_c, \mpi, m_K)$-space. For values of the chemical potential
above the chiral phase transition is of first-order while for values
below the surface the transition lies in the crossover region. With or
without anomaly the critical surface grows out perpendicular from the
$(\mpi,m_K)$-mass plane at $\mu = 0$. The tangent plane to the
critical surface has a decreasing slope for larger masses as
expected from Fig.~\ref{fig:phasediag_muB} or
Fig.~\ref{fig:phasediag_muB_mpiK}. Thus, this model study excludes the
so-called nonstandard scenario, found in a recent lattice analysis
with imaginary chemical potentials, where the first-order region
shrinks as the chemical potential is turned on
\cite{deForcrand:2006pv}. In the nonstandard scenario the bending of
the critical surface has the opposite sign and the physical realistic
mass point remains in the crossover region for any $\mu$. In the
Figure the physical mass point is denoted by an arrow.

Since the critical chemical potential $\mu_c$ cannot grow arbitrarily
the surface must have a boundary and hence stops to exist for larger
$(\mpi,m_K)$-masses which are not shown in the Figure.

In Ref.~\cite{Kovacs:2006ym} the \lsm with quarks in an one-loop
approximation, based on optimized perturbation theory, was evaluated
and an bending of the critical surface away from the $\mk$-axis at
$\mpi =0$ for a kaon masses greater than $400$ MeV was observed. The
precise value of $\mk$ for the onset of the bending depends on the
order of the used chiral perturbation theory ($\chi$PT) for the baryon
mass extrapolations. Thus, the unphysical bending indicates that the
validity range of the $\chi$PT for the baryons, used for the model
parameter extrapolation away from the physical point, was exceeded. As
a consequence, no (tri)critical point on the $\mk$-axis for $\mu=0$
where the boundary of the first-order transition region terminates,
can be located.

\begin{figure*}
  \subfigure[{$\ $} with $\ua$ anomaly]{\label{sfig:matteschnitt_muc}
    \includegraphics[width=0.45\linewidth]{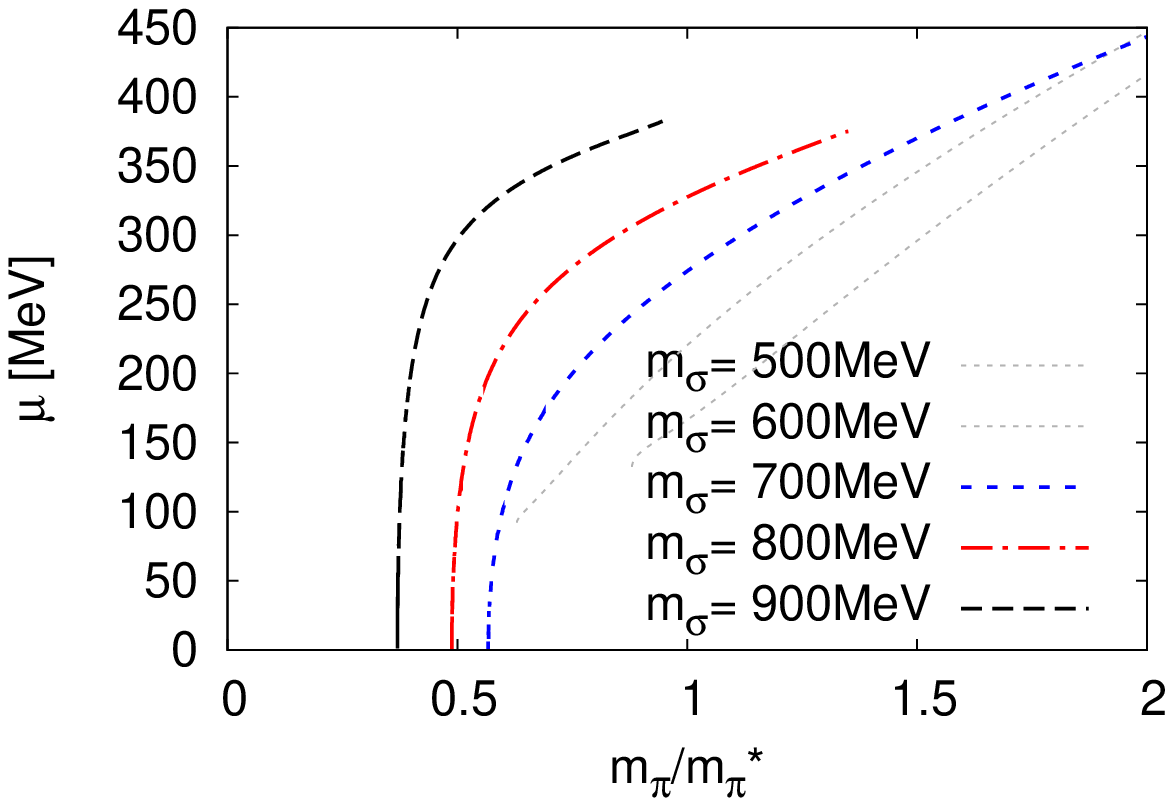}}
  \subfigure[{$\ $} without $\ua$
  anomaly]{\label{sfig:matteschnitt_c0_muc}
    \includegraphics[width=0.45\linewidth]{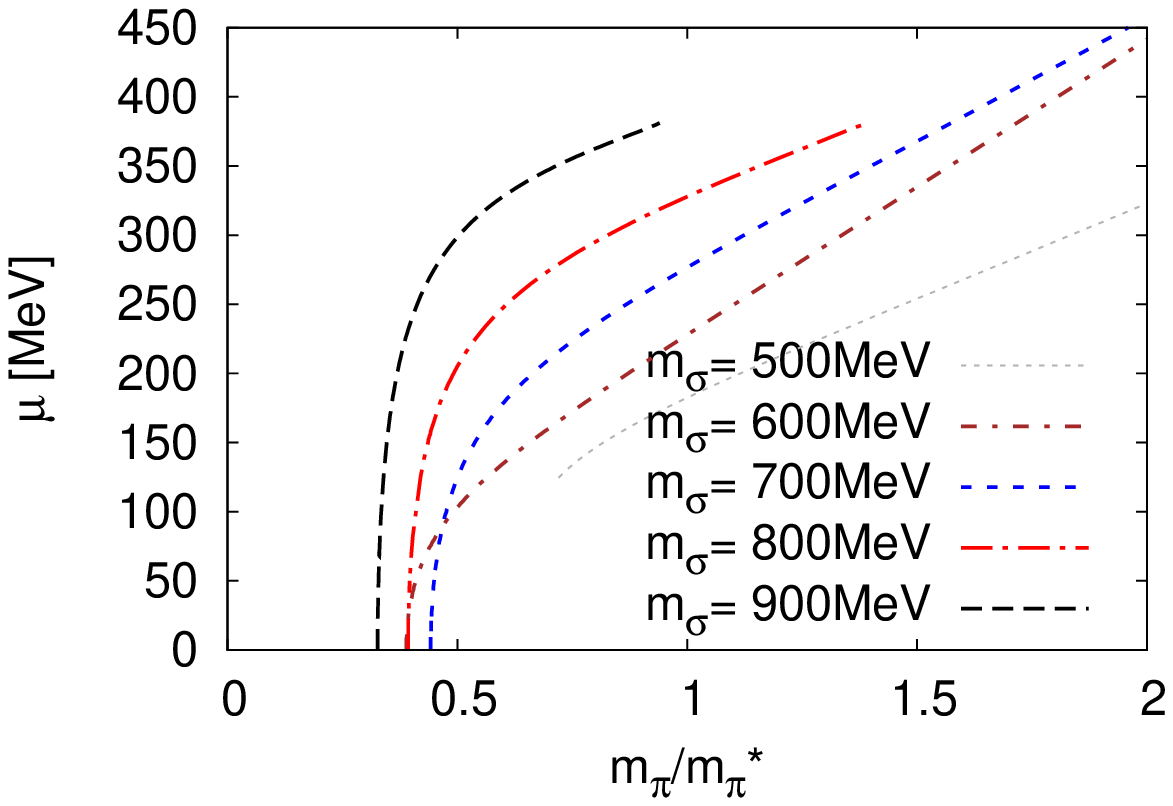}}
  \caption{Five cross sections of the chiral critical surface (left
    panel with, right panel without anomaly) for
    $\msig =500, \dots 900$ MeV. As trajectory in the ($\mpi, \mk$)
    plane we have chosen the path through the physical point towards
    the chiral limit, i.e.~we have kept the ratio of
    $m_{\pi}/m_{K} = m_{\pi}^{*}/m_{K}^{*}$ fixed. $m_{\pi}^{*}$ and
    $m_{K}^{*}$ denote the physical masses. See text for details.}
  \label{fig:matteschnitt}
\end{figure*}

In this work no strong bending of the surface away from the $\mk$-axis
is seen for kaon masses not larger than $600$ MeV but it also seems
that it never approaches the $\mk$-axis at least for $\msig
=800$ MeV. On the other hand, including the anomaly we find a critical
$\mpi^{\textrm crit} \sim 177$ MeV where the critical surface
intersects the solid line in the left panel of
\Fig{fig:columbia_pset3_1003}. On this line and for $\mpi
>\mpi^{\textrm crit}$ the transition turns into a smooth crossover.
On the contrary, without anomaly (right panel in the Figure), no critical
pion mass is found at $\mu=0$ at least for values below $200$ MeV.
This means that the phase transition on the solid line is still of
first-order similar to the findings of \cite{Lenaghan:2000kr} where
the influence of the anomaly on the phase transition for vanishing
chemical potentials within a $SU(3)\times SU(3)$ \lsm without fermions
in Hartree approximation has been investigated.

Furthermore, the effect of the $\ua$ anomaly on the shape of the
surface is rather marginal for a kaon mass greater than $400$ MeV
(cf.~both panels in \Fig{fig:columbia_pset3_1003}). For larger kaon
masses the strange sector decouples from the light sector and the
chiral phase transition is basically driven by the (light) nonstrange
particles.

On the other hand, for a kaon masses smaller than $400$ MeV we see a
considerable influence of the anomaly on the shape of the critical
surface. Without anomaly the region of first-order phase transitions
at $\mu=0$ is reduced which is in contrast to the results obtained
with a \lsm without quarks \cite{Lenaghan:2000kr}. In this reference,
it is found that the first-order transition region at $\mu=0$ grows
with and without anomaly for increasing sigma masses. For sigma masses
greater than $600$ MeV and without anomaly the physical point is well
located within the first-order region while it is always in the
crossover region with anomaly.

Including quarks we obtain an opposite tendency: the physical point is
always in the crossover region and for larger sigma masses the size of
the first-order transition region at $\mu=0$ decreases as can be seen
from \Fig{fig:matteschnitt}. In this figure five crosssections, for
$\msig = 500 \ldots 900$ MeV of the chiral critical surface with
(left) and without anomaly (right panel) are shown as a function of
the pion mass. As a trajectory in the $(\mpi, \mk)$-plane a path
through the physical point towards the chiral limit has been chosen
for these figures. This path is given by fixing the pion over kaon
mass ratio to the physical one, i.e. setting $\mk/\mpi = \mk^*/\mpi^*$
where the star denotes the corresponding physical masses.

For larger $\msig$ values the chiral critical surface with anomaly
moves to smaller pion masses. This effect is less pronounced if the
anomaly is neglected (right panel). In accordance with
\Fig{fig:phasediag_muB} the chiral critical surface line for $\msig =
900$ MeV terminates before the physical point ($\mpi/\mpi^* = 1$). Due
to the positive $m^2$ parameter we cannot evaluate the chiral critical
surface for arbitrary pion and kaon masses for smaller values of
$\msig$ as already mentioned. Nevertheless, the results for smaller
sigma masses are shown as dashed light curves in both panels and stop
at certain pion mass ratios.

\begin{figure*}
  \subfigure[{$\ $} with $\ua$ anomaly]{\label{sfig:matteschnitt_Tc}
    \includegraphics[width=0.45\linewidth]{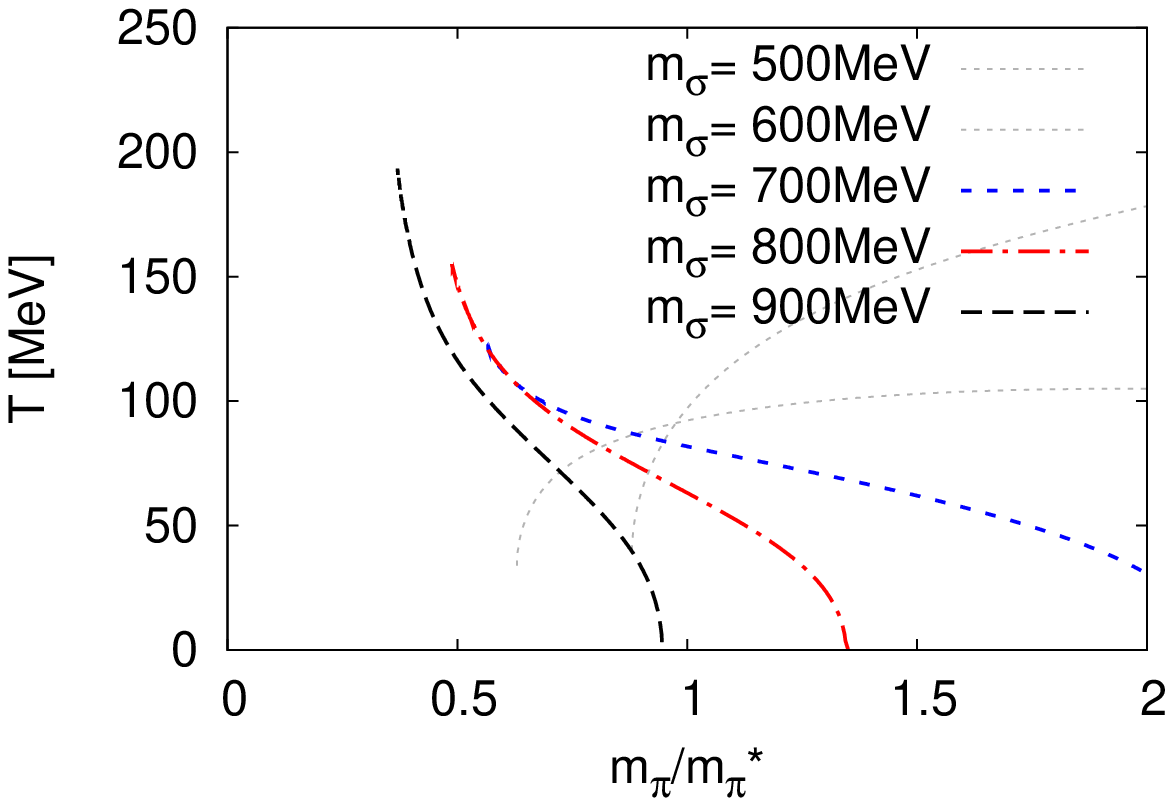}}
  \subfigure[{$\ $} without $\ua$
  anomaly]{\label{sfig:matteschnitt_c0_Tc}
    \includegraphics[width=0.45\linewidth]{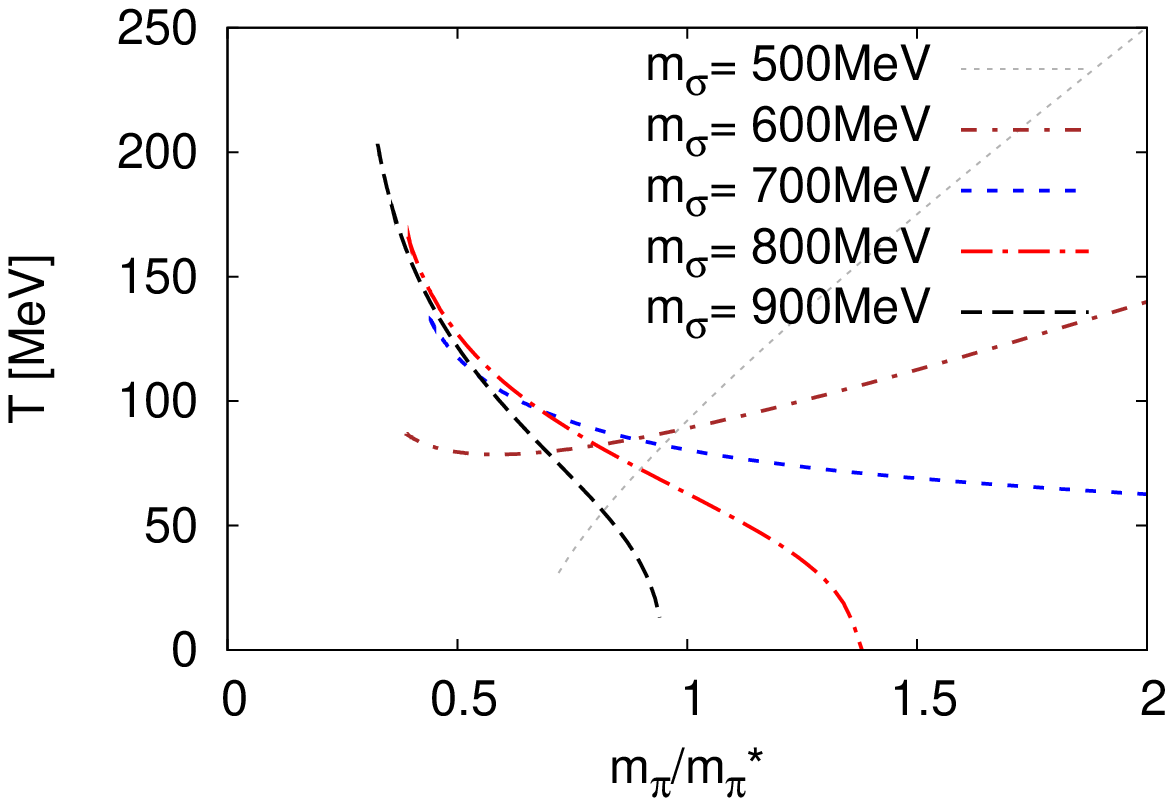}}
  \caption{Similar to Fig.~\ref{fig:matteschnitt} but the critical
    chemical potentials are replaced by the corresponding critical
    temperatures.}
  \label{fig:matteschnitt_T}
\end{figure*}

It is instructive to replace in the last \Fig{fig:matteschnitt} the
critical chemical potential with the critical temperature and
investigate its mass sensitivity. This leads to
\Fig{fig:matteschnitt_T} where the critical temperatures of the CEP's
as a function of the pion mass similar to \Fig{fig:matteschnitt} are
plotted for five different $\sigma$ meson masses with (left panel) and
without anomaly (right panel). In general, the curves start at zero
temperature and grow to a certain finite value for decreasing pion
masses. For the parameter sets without spontaneous symmetry breaking
the curves with anomaly for $\msig = 500$ and $600$ MeV, and without
anomaly for $\msig = 500$ MeV, show the opposite behavior. In this
case, i.e.~for smaller values of $\msig$, and large unphysical pion
masses $\mpi/\mpi^* > 1$, which is realized in lattice simulations,
the location of the CEP moves towards the $T$-axis. For decreasing pion
masses the critical temperatures decrease meaning that the CEP moves
towards the $\mu$-axis in the chiral limit. Consequently, for these
parameter sets no first-order transition occurs in the chiral limit as
already mentioned. Nevertheless, all of curves intersect roughly at
the physical point ($\mpi/\mpi^* =1$). Only the extrapolation towards
the chiral limit is questionable as can be seen by the positive slope
of these curves for decreasing pion masses.

\section{Summary}
\label{sec:summary}

In the present work chiral symmetry restoration in hot and dense
hadronic matter is analyzed. As an effective realization of the
spontaneous breaking of chiral symmetry in the vacuum, the $SU(3)_L
\times SU(3)_R$ symmetric \lsm with quarks has been used. Within this
model, the grand thermodynamic potential was evaluated in the
mean-field approximation. Six of the seven model parameters are
fixed to the low-lying pseudoscalar mass spectrum, which is
experimentally well-known and to the scalar sigma meson
mass. Since the experimental situation in the scalar sector is not very
certain we have varied the value of the sigma mass over a wide range and
have investigated its consequences for the physics. The remaining
model parameter, the Yukawa coupling, is determined by the non-strange
constituent quark mass while the condensates are governed by the PCAC
relation.

At the physical mass point for various sigma masses a smooth
finite temperature chiral crossover at zero density and a first-order
transition for finite chemical potential at zero temperature is found.
The $\ua$ anomaly has only little influence on the strange
condensate while no modification of the light condensate at the
physical point is observed. The pseudocritical crossover temperature
depends on the choice of the sigma meson mass and coincides with recent
lattice simulations for mass values in between $m_{\sigma} = 600
\ldots 800$ MeV.

In our approximation no negative squared meson masses are generated in
the medium as is the case for the \lsm without
quarks. Low-energy theorems like the Goldstone Theorem or the Ward
identities are not only fulfilled in the vacuum but also in the
medium. This enables a careful analysis of the chiral symmetry
restoration pattern of the meson nonets with and without axial
anomaly. An anticrossing of the scalar $\sigma$-$f_0(1370)$ masses at
higher temperature is seen which is reflected in the corresponding
mixing angle investigation.

For a sigma masses below $900$ MeV a CEP is found in the phase
diagram. In contrast to a similar calculation in the NJL model at this
point the sigma meson mass drops below the pion mass. This behavior is
analogous to the findings in the two flavor case.

The chiral critical surface in the $(m_{\pi},m_{K})$-plane always has a
positive slope such that the so-called non-standard scenario can be
excluded. In the chiral limit the expected first-order transition at
$\mu=0$ is found. A large modification caused by the anomaly is
visible. With anomaly the first-order region in the $\mu=0$ plane is
enlarged for smaller masses. For larger kaon masses the shape of the
critical surface becomes independent of the anomaly. Furthermore, for
increasing $\msig$ the surface becomes steeper.

One drawback of the effective model used here is the lack of confinement
properties which certainly modify the thermodynamics in the broken
phase. A first step towards an implementation of gluonic degrees of
freedom which could mimic certain confinement aspects can be achieved
by adding the Polyakov loop to the quark dynamics. Work in this
direction is in progress \cite{BJSMW2008}.

\subsection*{Acknowledgments}

MW was supported by the BMBF grant 06DA123 and he thanks the members
of the FWF-funded Doctoral Program ``Hadrons in vacuum, nuclei and
stars'' at the Institute of Physics of the University of Graz for
their hospitality and support. We are grateful to D.H.~Rischke and
J.~Wambach for a careful reading of the manuscript.

\appendix
\section{Parameter fits}
\label{app:parameters}

In this appendix several used parameter fits for the linear sigma
model (\lsm) with three quark flavors are collected.

As experimental input we have chosen the low-lying pseudoscalar meson
mass spectrum ($\mpi$, $\mk$, $\meta$), the constituent quark mass
$m_q$ and the pion and kaon decay constant. To be more precise, for
the fit with anomaly ($c\neq0$) the sum $\meta^2$+$\metap^2$ is chosen
as input while for $c=0$ the $\meta$ has not been used as input. The
experimental values, taken from Ref.~\cite{Yao:2006px}, are listed in
the last line of \Tab{tab:massesvac} for comparison. Since the chiral
$\sigma$-particle is a broad resonance its mass is not known
precisely. We therefore have used different input values for $\msig$
in the range of $400-1000$ MeV and have refitted the remaining
parameters of the model accordingly. It is remarkable that larger
sigma meson masses are not adjustable since $\msig$ as a function of
the quartic coupling $\lambda_1$ saturates around 1100 MeV.

\begin{table*}
\begin{tabular}{|c||c|c||d|d|d|d|d|d|d|}
\hline  \mwmc{|c||}{$\msig$} & \mwc{$\mpi$} & \mwc{$\mk$} & \mwc{$\metap$} & \mwc{$\meta$} & \mwc{$\thetap$} & \mwc{$m_{a_{0}}$} & \mwc{$\mkappa$} & \mwc{$\mf0$} & \mwc{$\thetas$}  \\\hline
\T 400 &138 &496 &138.0 &634.8 &35.3 &850.4 &1124.3  &1257.3 &16.7\\
\T 500 &138 &496 &138.0 &634.8 &35.3 &850.4 &1124.3  &1267.4 &18.7\\
\T 600 &138 &496 &138.0 &634.8 &35.3 &850.4 &1124.3  &1282.3 &21.5\\
\T 700 &138 &496 &138.0 &634.8 &35.3 &850.4 &1124.3  &1304.9 &25.5\\
\T  800 &138 &496 &138.0 &634.8 &35.3 &850.4 &1124.3  &1341.4 &31.3\\
\T 900 &138 &496 &138.0 &634.8 &35.3 &850.4 &1124.3 &1408.0 &40.0\\
\T \B 1000 &138 &496 &138.0 &634.8 &35.3 &850.4 &1124.3 &1563.4 &53.2\\
\hline \hline
\T  400 &138 &496 &963.0 &539.0 &-5.0 &1028.7 &1124.3 &1198.4 &14.9  \\
\T  500 &138 &496 &963.0 &539.0 &-5.0 &1028.7 &1124.3 &1207.5 &16.9  \\
\T  600 &138 &496 &963.0 &539.0 &-5.0 &1028.7 &1124.3 &1221.1 &19.9  \\
\T  700 &138 &496 &963.0 &539.0 &-5.0 &1028.7 &1124.3 &1242.3 &24.2  \\
\T  800 &138 &496 &963.0 &539.0 &-5.0 &1028.7 &1124.3  &1278.0 &30.7  \\
\T 900 &138 &496 &963.0 &539.0 &-5.0 &1028.7 &1124.3  &1348.0 &40.9  \\
\T \B 1000 &138 &496 &963.0 &539.0 &-5.0 &1028.7 &1124.3  &1545.6 &57.1 \\
\hline\hline
\mwmc{|c||}{\T  400-1200} & 138.0 & 496 & 957.78 & 547.5 & &984.7 & 1414  & \mwc{1200-1500} & \\
\hline
\end{tabular}
\caption{Meson masses and mixing angles in the
  vacuum for different sets of parameters. The first six columns are
  input while the remaining columns are predictions. Upper block:
  without ($c=0$) $\ua$ anomaly, lower block: with $\ua$ anomaly. Last line:
  experimental values from the PDG \cite{Yao:2006px}.}
  \label{tab:massesvac}
\end{table*}


In \Tab{tab:massesvac} all resulting meson masses are listed. The
upper block contains the fit without anomaly and the lower block the
fit including the anomaly. Except for the pseudoscalar masses
the first four/five columns in the table respectively, all other (scalar) masses
and the mixing angles are predictions of the model.


In \Tab{tab:psets} the values for the six mesonic model parameters of
the \lsm with and without explicit $\ua$ symmetry breaking are
summarized. The Yukawa coupling is always kept fixed to $g \sim 6.5$
corresponding to a constituent quark mass of $m_q = 300$ MeV. For this
Yukawa coupling the strange constituent quark mass is predicted to be
$m_s \approx 433$ MeV. The decay constants, $\fpi = 92.4$ MeV and $\fk
= 113$ MeV, are also kept constant for all fits. It is interesting to
realize that for small values of $\msig$ and with $\ua$ symmetry
breaking the mass parameter $m^2$ changes sign and becomes positive
when fitted to realistic masses. As a consequence spontaneous symmetry
breaking is lost in the chiral limit and all condensates will vanish
in this limit. This happens for $\msig \leq 700$ MeV. Even without
anomaly breaking ($c=0$) a similar phenomenon can be seen. For this
case the masses are smaller when this case sets in, i.e.~$\msig \leq
500$ MeV. This is the motivation for our choice of $\msig =800$ MeV.
For this parameter set we can investigate the mass sensitivity of the
chiral phase transition over arbitrarily explicit symmetry breaking
values including the chiral limit. The choice $\msig = 800$ MeV for
the parameter fit without anomaly is also in agreement with
\cite{Lenaghan:2000ey} where $\msig=600$ MeV is a misprint in this
reference. For larger $\msig$ values the quartic coupling $\lambda_1$
increases significantly which restricts the parameter fits for $\msig$
larger than $1000$ MeV.

\begin{table*}
\begin{tabular}{| d || d | d | d | d | d | d |}
\hline
\multicolumn{1}{|c||}{\T \B $m_\sigma \text{[MeV]}$}& \mwc{$c \text{[MeV]}$} & \mwc{$\lambda_1$}  	& \mwc{$m^2 [\MeV^2]$} & \mwc{$\lambda_{2}$} 	& \mwc{$h_{x}[\MeV^3]$} & \mwc{$h_{y} [\MeV^3]$}\\
\hline
\T 400 & 0	& -24.55 	&  +(309.41)^2	& 82.47 	& (120.73)^{3} & (336.41)^{3}\\
\T 500	& 0	& -21.24 	&  +(194.82)^2	& 82.47 	& (120.73)^{3} & (336.41)^{3}\\
\T 600 	& 0	& -17.01 	& -(189.85)^2	& 82.47 	& (120.73)^{3} & (336.41)^{3}\\
\T 700 	& 0	& -11.61 	& -(360.91)^2	& 82.47 	& (120.73)^{3} & (336.41)^{3}\\
\T 800 	& 0 	&  -4.55	& -(503.55)^2	& 82.47	& (120.73)^{3} & (336.41)^{3}\\
\T 900  & 0	& 5.56	 	& -(655.82)^2	& 82.47 	& (120.73)^{3} & (336.41)^{3}\\
\T \B 1000	& 0	& 24.22 & -(869.50)^2	& 82.47 	& (120.73)^{3} & (336.41)^{3}\\
\hline
\T 400 	& 4807.84 & -5.90	& +(494.75)^2	& 46.48	& (120.73)^{3} & (336.41)^{3}\\
\T 500 	& 4807.84 & -2.70	& +(434.56)^2	& 46.48	& (120.73)^{3} & (336.41)^{3}\\
\T 600 	& 4807.84 &  1.40	& +(342.52)^2	& 46.48	& (120.73)^{3} & (336.41)^{3}\\
\T 700	& 4807.84 &  6.62	& +(161.98)^2	& 46.48	& (120.73)^{3} & (336.41)^{3}\\
\T 800 	& 4807.84 & 13.49  	& -(306.26)^2	& 46.48	& (120.73)^{3} & (336.41)^{3}\\
\T 900 	& 4807.84 & 23.65  	& -(520.80)^2	& 46.48	& (120.73)^{3} & (336.41)^{3}\\
\T \B 1000	& 4807.84 & 45.43	& -(807.16)^2	& 46.48	& (120.73)^{3} & (336.41)^{3}\\
\hline
\end{tabular}

\caption{
  Different parameter sets for various $m_\sigma$ with ($c\neq 0$) and
  without ($c=0$) $\ua$ anomaly. }
  \label{tab:psets}
\end{table*}

\section{Meson masses}
\label{sec:meson_masses}

The scalar $J^P=0^+$ and pseudoscalar $J^P=0^-$ meson masses are
defined by the second derivative w.r.t. the corresponding scalar and
pseudoscalar fields $\varphi_{s,a} = \sigma_a$ and
$\varphi_{p,a} = \pi_a \ (a=0,\ldots,8)$ of the grand
potential $\Omega (T,\mu_f)$, \Eq{eq:grand_pot}, evaluated at its
minimum. The minimum is given by vanishing expectation values of all
scalar and pseudoscalar fields but only two of them, $\bar \sigma_x$
and $\bar \sigma_y$, are nonzero.
\begin{equation}
  \label{eq:meson_masses}
  m^2_{i,{ab}} = \left. \frac{\partial^2 \Omega
      (T,\mu_f )}{\partial \varphi_{i,a} \partial \varphi_{i,b}}
    \right\vert_{\rm min}\quad;\quad i=s, p\ .
\end{equation}
In the vacuum the contribution from the quark potential vanishes.
Hence, only the mesonic part of the potential determines the mass
matrix completely. The squared mass matrix is diagonal and due to
isospin $SU(2)$ symmetry several matrix entries are degenerate. We
begin with the scalar, $J^P=0^+$, sector corresponding to $i=s$. The
squared mass of the $a_0$ meson is given by the (11) element which is
degenerate with the (22) and (33) elements. Similar, the squared
$\kappa$ meson mass is given by the (44) element which is also
degenerated with the (55), (66) and (77) elements. The $\sigma$ and
$f_0(1370)$ meson masses are obtained by diagonalizing the (00)-(88)
sector of the mass matrix introducing in this way a mixing angle
$\thetas$. Explicitly, the squared masses for scalar sector,
formulated in the nonstrange-strange basis, are
\begin{widetext}
\begin{align}
  m^2_{a_0} &= m^2 + \lambda_1 (\bsig_x^2 + \bsig_y^2) + \frac{3
    \lambda_2}{2} \bsig_x^2 +\frac{\sqrt{2} c}{2} \bsig_y\ ,\\
  m^2_{\kappa} &= m^2 + \lambda_1 (\bsig_x^2 + \bsig_y^2) +
  \frac{\lambda_2}{2} \left(\bsig_x^2 + \sqrt{2} \bsig_x \bsig_y +
    2 \bsig_y^2\right) + \frac{c}{2}\bsig_x\ ,\\[1ex]
  m^2_\sigma &= m^2_{s,{00}} \cos^{2}\thetas +
  m^2_{s,{88}}\sin^{2}\thetas + 2  m^2_{s,{08}} \sin \thetas \cos
  \thetas\ ,\\[1ex]
  m^2_{f_0}  &= m^2_{s,{00}} \sin^{2}\thetas +
  m^2_{s,{88}}\cos^{2}\thetas - 2  m^2_{s,{08}} \sin \thetas \cos \thetas\\[1ex]
\text{with}\quad   m^2_{s,{00}} &= m^2 + \frac{\lambda_1}{3} (7 \bsig_x^2 + 4
  \sqrt{2}\bsig_{x}\bsig_{y} + 5 \bsig_y^2) + \lambda_2(\bsig_x^2 +
  \bsig_y^2) -\frac{\sqrt{2}c}{3}(\sqrt{2} \bsig_x +
  \bsig_y)\ ,\nonumber \\
 m^2_{s,{88}} &= m^2 + \frac{\lambda_1}{3} (5 \bsig_x^2 - 4
  \sqrt{2}\bsig_{x}\bsig_{y} + 7 \bsig_y^2) +
  \lambda_2(\frac{\bsig_x^2}{2} + 2\bsig_y^2)
  +\frac{\sqrt{2}c}{3}\left(\sqrt{2} \bsig_x -
  \frac{\bsig_y}{2}\right)\ ,\nonumber  \\
  m^2_{s,{08}} &= \frac{2 \lambda_1}{3}\left(\sqrt{2}\bsig_x^2 - \bsig_x \bsig_y - \sqrt{2}\bsig_y^2 \right)
  + \sqrt{2}\lambda_2 \left(\frac{\bsig_x^2}{2} - \bsig_y^2 \right)
  + \frac{c}{3\sqrt{2}} \left(\bsig_x - \sqrt{2}\bsig_y\right)
	\end{align}
\end{widetext}

The situation for the pseudoscalar sector ($i=p$) is completely
analogous with the following labeling: the squared pion mass is
identified with the (11) element and the squared kaon mass with the
(44) element of the pseudoscalar mass matrix. Similar to the scalar
case, the $\eta$ and $\eta'$ mass are obtained by diagonalizing the
(00)-(88) sector and accordingly a pseudoscalar mixing angle
$\thetap$ is introduced. Explicitly, the squared masses for the
pseudoscalar sector are
\begin{widetext}
\begin{align}
  m^2_\pi &= m^2 + \lambda_1 (\bsig_x^2 + \bsig_y^2) +
  \frac{\lambda_2}{2} \bsig_x^2 -\frac{\sqrt{2} c}{2} \bsig_y\\
  m^2_K &= m^2 + \lambda_1 (\bsig_x^2 + \bsig_y^2) +
  \frac{\lambda_2}{2} \left(\bsig_x^2 - \sqrt{2} \bsig_x \bsig_y +
    2 \bsig_y^2\right) - \frac{c}{2}\bsig_x\\[1ex]
    m^2_{\eta'} &= m^2_{p,{00}} \cos^{2}\thetap + m^2_{p,{88}}\sin^{2}\thetap + 2 m^2_{p,{08}} \sin \thetap \cos \thetap\\[1ex]
  m^2_{\eta} &= m^2_{p,{00}} \sin^{2}\thetap + m^2_{p,{88}}\cos^{2}\thetap - 2 m^2_{p,{08}} \sin \thetap \cos \thetap\\[1ex]
\text{with}\quad    m^2_{p,{00}} &= m^2 + \lambda_1 (\bsig_x^2 + \bsig_y^2)
  + \frac{\lambda_2}{3}(\bsig_x^2 + \bsig_y^2) +\frac{c}{3}(2
  \bsig_x + \sqrt{2} \bsig_y)\nonumber \\
  m^2_{p,{88}} &= m^2 + \lambda_1 (\bsig_x^2 + \bsig_y^2)
  + \frac{\lambda_2}{6}(\bsig_x^2 + 4 \bsig_y^2) -\frac{c}{6}(4
  \bsig_x - \sqrt{2}\bsig_y)\nonumber \\
  m^2_{p,{08}} &= \frac{\sqrt{2} \lambda_2}{6} (\bsig_x^2 - 2
  \bsig_y^2) - \frac{c}{6}(\sqrt{2}\bsig_x - 2 \bsig_y)\nonumber
\end{align}
\end{widetext}
Both mixing angles are given by
\begin{equation}
  \tan 2\Theta_i = \frac{2 m^2_{i,{08}}}{m^2_{i,{00}}-m^2_{i,{88}}}\
  ,\  i=s,p\ .
\end{equation}

In the medium the meson masses are further modified by the quark
contribution (\ref{eq:quark_pot}). In order to evaluate the second
derivate (\ref{eq:meson_masses}) for the quark contribution the
complete dependence of all scalar and pseudoscalar meson fields,
cf.~\Eq{eq:fields}, in the quark masses has to be taken into account.
The resulting quark mass matrix can be diagonalized. Finally, we
obtain the expression
\begin{widetext}
 \begin{equation}
   \left.\frac{\partial^2 \Omega_{\bar{q}q}(T,\mu_f)}{\partial
      \varphi_{i,\alpha}
    \partial \varphi_{i,\beta}}\right\vert_{\rm min} = \nu_{c}\sum_{f=q,s} \int
  \frac{d^{3}p}{(2\pi)^{3}} \frac{1}{2E_{q,f}} \left[\left(n_{q,f} + n_{\bar{q},f}
  \right) \left(m^2_{f,\alpha \beta} - \frac{m^2_{f,\alpha} m^2_{f,\beta}}{2 E^{2}_{q,f}}\right)
  -\frac{\left( b_{q,f}+ b_{\bar{q},f}\right)}
 {2 E_{q,f}T} m^2_{f,\alpha} m^2_{f,\beta}\right]\\
\end{equation}
\end{widetext}
where we have introduced the short hand notation
$m^2_{f,a} \equiv \partial m^2_f/\partial \varphi_{i,a}$ for the quark
mass derivative w.r.t.~the meson fields $\varphi_{i,a}$, the quark
function
\begin{equation}
b_{q,f}(T,\mu_f)= n_{q,f}(T,\mu_f) (1-n_{q,f}(T,\mu_f))
\end{equation}
and correspondingly the antiquark function
$b_{\bar{q},f}(T,\mu_f) = b_{q,f}(T,-\mu_f)$. The index $i$
distinguishes between a scalar and pseudoscalar field which we omit in
the following. In \Tab{tab:mmesonderivs} all second quark mass
derivatives w.r.t.~the meson fields replaced by the non-vanishing
vacuum expectation values in the nonstrange-strange basis are
collected. Despite the $SU(2)$ isospin symmetry the derivatives are
different for the up- and down-quark sector. In the table
\ref{tab:mmesonderivs}, left column block, the sum over the two light
quark flavors is shown which leads to large cancellations.
\begin{table}
\newcolumntype{C}{>{$}c<{$}}
\begin{tabular}{|CC||CC|CC|}\hline
 &  & \T \B m_{l,\alpha}^2 m_{l,\beta}^2/g^{4} & m_{l,\alpha\beta}^2/g^{2} & m_{s,\alpha}^2 m_{s,\beta}^2/g^{4} & m_{s,\alpha\beta}^2/g^{2} \\
\hline
\T \B \bsig_{0} & \bsig_{0} & \frac{1}{3}\bsig_x^2 & \frac{2}{3} & \frac{1}{3}\bsig_y^2 &  \frac{1}{3}\\
\T \B \bsig_{1} & \bsig_{1} & \frac{1}{2}\bsig_x^2 & 1 & 0 & 0\\
\T \B \bsig_{4} & \bsig_{4} & 0 & \bsig_{x}\frac{\bsig_{x} + \sqrt{2}\bsig_{y}}{\bsig_x^2-2\bsig_y^2} & 0 & \bsig_{y}\frac{\sqrt{2}\bsig_{x} + 2 \bsig_{y}}{2\bsig_y^2-\bsig_x^2}\\
\T \B \bsig_{8} & \bsig_{8} & \frac{1}{6}\bsig_{x}^{2} & \frac{1}{3} & \frac{2}{3}\bsig_{y}^{2} & \frac{2}{3}\\
\T \B \bsig_{0} & \bsig_{8} & \frac{\sqrt{2}}{6} \bsig_x^2& \frac{\sqrt{2}}{3} & -\frac{\sqrt{2}}{3}\bsig_{y}^{2} & -\frac{\sqrt{2}}{3}\\
\hline
\hline
\T \B \pi_{0} & \pi_{0} & 0 & \frac{2}{3} & 0 & \frac{1}{3}\\
\T \B \pi_{1} & \pi_{1} & 0 & 1 & 0 & 0\\
\T \B \pi_{4} & \pi_{4} & 0 &  \bsig_{x}\frac{ \bsig_{x} - \sqrt{2}\bsig_{y}}{\bsig_{x}^{2} - 2\bsig_{y}^{2}} & 0 &  \bsig_{y}\frac{ \sqrt{2}\bsig_{x} - 2\bsig_{y}}{\bsig_{x}^{2}-2\bsig_{y}^{2}}\\
\T \B \pi_{8} & \pi_{8} & 0 & \frac{1}{3} & 0 & \frac{2}{3}\\
\T \B \pi_{0} & \pi_{8} & 0 & \frac{\sqrt{2}}{3}& 0 & -\frac{\sqrt{2}}{3}\\\hline
\end{tabular}
\caption{Squared quark mass second  derivatives with respect  to the
  meson fields  evaluated at the  minimum. Left column block, the sum
  over two light quark flavors, denoted by index $l$ and right column
  block only the strange quark flavor, index $s$.}
  \label{tab:mmesonderivs}
\end{table}

\section{Isoscalar mixing}
\label{sec:eta_mixing} %

In this appendix a collection of relations describing the mixing of
isoscalar states in the pseudoscalar and scalar multiplet is
presented. The isoscalar ($I=0$) pseudoscalar states in the
octet-singlet ($\eta_8,\eta_0$) basis are defined by
\begin{equation}
  \ket{\eta_8} = \frac{ 1}{\sqrt{6}}\ket{u\bar u + d \bar d -2 s \bar
    s},\quad
  \ket{\eta_0} = \frac{ 1}{\sqrt{3}}\ket{u\bar u + d \bar d + s \bar s}.
\end{equation}
For a realistic flavor breaking in the vacuum the physical $\eta$
meson is close to the $\eta_8$ and $\eta'$ to $\eta_0$.

The eigenstates in the flavor nonstrange-strange ($\ens,\es$) basis
are given by
\begin{equation}
  \ket{\ens} = \frac{ 1}{\sqrt{2}}\ket{u\bar u + d \bar d },\
  \ket{\es} = \ket{s \bar s}\ .
\end{equation}
These states are associated to each other by a rotation with an angle
$\alpha = - \arctan\sqrt{2} \sim -54.74^\circ$
\begin{equation}
\left( \begin{array}{l}
\ket{\ens} \\
\ket{\es}
\end{array}
\right) = \frac{1}{\sqrt{3}}\left( \begin{array}{rr}
1 &  \sqrt{2} \\
-\sqrt{2} &  1
\end{array}
\right) \left( \begin{array}{l}
\ket{\eta_8}\\
\ket{\eta_0}
\end{array}\right)\ .
\end{equation}

Diagonalization of the mass matrix in the ($\eta_8,\eta_0$) basis is
achieved by the introduction of the pseudoscalar mixing angle
$\thetap$. This yields the relations
\begin{equation}
\left( \begin{array}{l}
\ket{\eta} \\
\ket{\eta'}
\end{array}
\right) = \left( \begin{array}{lr}
\cos \thetap &  - \sin \thetap \\
\sin \thetap &  \cos \thetap
\end{array}
\right) \left( \begin{array}{l}
\ket{\eta_8}\\
\ket{\eta_0}
\end{array}\right)\ .
\end{equation}
For the ($\ens,\es$) basis a similar relation with the mixing angle
$\phi_p$ holds
\begin{equation}
\left( \begin{array}{l}
\ket{\eta} \\
\ket{\eta'}
\end{array}
\right) = \left( \begin{array}{lr}
\cos \phi_{p} &  - \sin \phi_{p} \\
\sin \phi_{p} &  \cos \phi_{p}
\end{array}
\right) \left( \begin{array}{l}
\ket{\ens} \\
\ket{\es}
\end{array}\right)\ .
\end{equation}
For vanishing mixing angle $\phi_p$ corresponding to $\thetap
=-\arctan \sqrt{2} \sim -54,7^\circ$ the $\eta$ tends to a pure
nonstrange $\ens$ and $\eta'$ to a pure strange $\es$. In contrast,
for a mixing angle $\phi_p = 90^\circ$ ($\thetap \sim +35,3^\circ$)
the ordering is reversed and $\eta \to \es$ and $\eta' \to \ens$. The
ordering transition occurs at $\phi_p = 45^\circ$ ($\thetap \sim
-9.74^\circ$).

The diagonalization of the mass matrix in the ($\ens,\es$) basis leads to
the masses
\begin{eqnarray}
  \label{eq:eta1}
  m^2_\eta \!\!\!&=&\!\!\! m^2_\ens \cos^2 \phi_{p} + m^2_\es \sin^2 \phi_{p} -
  m^2_{\es,\ens} \sin^2 (2\phi_{p}),\qquad \\
  \label{eq:etap1}
  m^2_{\eta'} \!\!\!&=&\!\!\! m^2_\ens \sin^2 \phi_{p} + m^2_\es \cos^2 \phi_{p} +
  m^2_{\es,\ens} \sin^2 (2\phi_{p}).
\end{eqnarray}
and to the mixing angle $\phi_p$ given by
\begin{equation}
\label{eq:etaphi}
  \tan 2\phi_p = \frac{2 m^2_{\es,\ens}}{m^2_{\es}-m^2_{\ens}}\ .
\end{equation}
Equivalently, these expressions can be rewritten in a form which do
not contain the mixing angle explicitly
\begin{eqnarray}
  \label{eq:etadiag}
  m^2_{\eta'} &=& \frac{ 1}{2} (m^2_\ens + m^2_\es +
  \Delta_{\ens,\es})\ ,\\
  \label{eq:etadiag2}
  m^2_{\eta} &=& \frac{ 1}{2} (m^2_\ens + m^2_\es -
  \Delta_{\ens,\es})\ ,
\end{eqnarray}
with
$\Delta_{\ens,\es} \equiv \sqrt{\left(m^2_\ens - m^2_\es\right)^2 + 4
  m^2_{\ens,\es}}$. Note, that these expressions are numerically more
stable compared to (\ref{eq:eta1} - \ref{eq:etap1}) because possible
ambiguities in the tangent (\ref{eq:etaphi}) do not appear here.

The matrix elements in the ($\ens,\es$) system are obtained by a base
change from the ones in the ($\eta_8,\eta_0$) basis with the result
\begin{eqnarray}
  m^2_\ens &=& \frac{ 1}{3} (2 m^2_{p,00} +m^2_{p,88}+2\sqrt{2}
  m^2_{p,08})\ , \nonumber \\
  m^2_{\es} &=& \frac{ 1}{3} (m^2_{p,00} +2
  m^2_{p,88}-2\sqrt{2} m^2_{p,08})\ , \\
  m^2_{\es,\ens} &=& \frac{ 1}{3} (\sqrt{2} (m^2_{p,00} -
  m^2_{p,88})- m^2_{p,08})\ .\nonumber
\end{eqnarray}

As a consequence,  the mixing angles $\phi_p$ and $\thetap$ are related by
\begin{equation}
  \phi_p = \thetap + \arctan\sqrt{2} \sim \thetap + 54.74^\circ \ .
\end{equation}

Furthermore, supposing $\meta \leq \metap$ one finds with
(\ref{eq:eta1}),(\ref{eq:etap1}) for $\phi_p \leq 45^\circ$
($\thetap \leq -9.74^\circ$) $m_\ens \leq m_\es$ while for
$\phi_p > 45^\circ$ the ordering of the masses in the
nonstrange-strange system are reversed.

Scalar mesons differ from the pseudoscalar ones only in the orbital
excitation. Hence, all quoted relations can be immediately converted
to the scalar ($\sigma$, $f_0$) complex with the corresponding
replacements, e.g.

\begin{equation}
\left( \begin{array}{l}
\ket{f_{0}} \\
\ket{\sigma}
\end{array}
\right) = \left( \begin{array}{lr}
\cos \phi_{s} &  - \sin \phi_{s} \\
\sin \phi_{s} &  \cos \phi_{s}
\end{array}
\right) \left( \begin{array}{l}
\ket{\signs} \\
\ket{\sigs}
\end{array}\right)\ .
\end{equation}

For an ideal scalar mixing angle $\phi_s = 90^\circ$ the $\sigma$
meson is a pure nonstrange state $\signs$ and $f_0 \to -\sigs$.
Furthermore, $\sigma$ matches with $\eta'$ and $f_0$ with $\eta$.

Since the mass of the $f_0$ meson is larger than $\msig$ we obtain for
an ideal mixing $\phi_s =90^\circ$ the ordering $m_\sigs > m_\signs$.

\end{document}